%% file: Safegaurding_IoT_Final_Files.tex
\documentclass[10pt, final, twocolumn, twoside, romanappendices]{IEEEtran}
\usepackage{amsthm,amssymb,graphicx,multirow,amsmath,color,amsfonts}
\usepackage[update,prepend]{epstopdf}
\usepackage[noadjust]{cite}
\usepackage[latin1]{inputenc}
\usepackage{bbm} 
\usepackage{pdfpages}
\usepackage{tabulary}
\usepackage{multirow}
\usepackage{paralist}
\usepackage{comment}
\usepackage{textcomp}
\usepackage{subcaption}
\usepackage{soul}
\usepackage{xcolor}
\usepackage{adjustbox}

\usepackage{dsfont}
\include{notation}

\input{defs_tikzpgf}

\usepackage{setspace}    

\begin{document}
\graphicspath{{./Figures/}}
\title{Safeguarding the IoT from Malware Epidemics: A Percolation Theory Approach}
\author{Ainur Zhaikhan, Mustafa A. Kishk, Hesham ElSawy, and Mohamed-Slim Alouini
\thanks{A. Zhaikhan, M. A. Kishk, and M.-S. Alouini are with King Abdullah University of Science and Technology (KAUST), Thuwal 23955-6900, Saudi Arabia. (emails: ainur.zhaikhan, mustafa.kishk, slim.alouini@kaust.edu.sa)}
	\thanks{H. ElSawy is with the Electrical Engineering Department, King Fahd University of Petroleum \& Minerals (KFUPM), Dhahran, 31261, Saudi Arabia (email: hesham.elsawy@kfupm.edu.sa). H. ElSawy acknowledges the support received from the deanship of scientific research (DSR) at KFUPM under grant no. DF191052.}
\vspace{-5mm}}

\maketitle 

\begin{abstract}

The upcoming Internet of things (IoT) is foreseen to encompass massive numbers of connected devices, smart objects, and cyber-physical systems. Due to the large-scale and massive deployment of devices, it is deemed infeasible to safeguard $100\%$ of the devices with state-of-the-art security countermeasures. Hence, large-scale IoT has inevitable loopholes for network intrusion and malware infiltration. Even worse, exploiting the high density of devices and direct wireless connectivity, malware infection can stealthily propagate through susceptible (i.e., unsecured) devices and form an epidemic outbreak without being noticed to security administration. A malware outbreak enables adversaries to compromise large population of devices, which can be exploited to launch versatile cyber and physical malicious attacks. In this context, we utilize {\em  spatial firewalls}, to safeguard the IoT from malware outbreak. In particular, spatial firewalls are computationally capable devices equipped with state-of-the-art security and anti-malware programs that are spatially deployed across the network to filter the wireless traffic in order to detect and thwart malware propagation. Using tools from percolation theory, we prove that there exists a critical density of spatial firewalls beyond which malware outbreak is impossible. This, in turns, safeguards the IoT from malware epidemics regardless of the infection/treatment rates. To this end, a tractable upper bound for the critical density of spatial firewalls is obtained. Furthermore, we characterize the relative communications ranges of the spatial firewalls and IoT devices to ensure secure network connectivity. The percentage of devices secured by the firewalls is also characterized.

\end{abstract}

\section{Introduction} \label{sec:intro}

The surging Internet of Things (IoT) and cyber physical systems (CPS) are extending wireless connectivity to billions of new devices of multitude heterogeneity~\cite{IoT_survey}. In addition to phones, tablets, and laptops, the IoT and CPS integrate appliances, sensors, actuators, machines, robots, vehicles, and many other smart objects to the wireless infrastructure. It is speculated that the numbers of IoT devices per square kilometers will be in the order of millions~\cite{million_IoT}. Such ubiquitous, large-scale, diverse, and  massive wireless connectivity is essential for big data aggregation and smart world automation, which is expected to improve almost every aspect in our lives~\cite{IoT_survey}. For instance, health care providers can access real time vital signals for patients through connected body sensors, which improves diagnostics, enables early disease detection, and decreases infection risks. Smart power grids utilize wireless connectivity of smart meters and field devices to improve energy generation and distribution. Intelligent transportation systems with connected/autonomous vehicles exploit wireless connectivity to improve road safety and reduce traffic congestion. Large-scale massive connectivity is also foundational for process automation in the next industrial revolution (i.e., industry 4.0). In addition to the aforementioned examples, IoT/CPS can bring unlimited potentials to many other verticals such as crowd management, public safety, agriculture, retail, etc.

The aforementioned benefits of IoT/CPS come at the cost of a plenty of new and challenging security threats~\cite{CPS_Sec_Survey,7938303,7906491,8437135,8166791}. The IoT and CPS devices are mainly installed and controlled via consumers who have limited knowledge about security threats and countermeasures. The imposed high competition between IoT vendors leads to overlooking cybersecurity aspects in order to accelerate the production of devices and reduce their prices. Furthermore, many of the IoT and CPS devices do not have sufficient energy, storage, or computational power to implement up-to-date anti-malware programs and/or sophisticated intrusion defense mechanisms~\cite{jumping, Epidemics1, Junaid, IoT}. In large-scale IoT/CPS networks, there is no distinct boundary between secured and public (i.e., unsecured) domains to enforce security policies on the incoming/outgoing traffic. The lack of per-device defense mechanisms and network-wide security administration open several loopholes for network intrusion and malware infiltration. Even worse, exploiting the high spatial density of devices and direct wireless connectivity (e.g., machine-to-machine and device-to-device communications), the malware infection can stealthily propagate from one device to another and form an {\em epidemic outbreak} without being noticed to security administration~\cite{jumping, wireless_worm, Nature_worm, Vehicular_epidemics}. Malware diffusion through the devices can be further accelerated via emerging beyond 5G technologies such as non-orthogonal multiple access (NOMA) and ultra-reliable low latency communications (URLLC), which are meant to enhance information dissemination. 

A malware outbreak gives adversaries the opportunity to compromise large population of IoT/CPS devices, which can then be used to launch versatile criminal and hostile attacks. Examples of generic IoT/CPS attacks are network-jamming, colluded eavesdropping, spoofing, denial of service, and data falsification~\cite{Broadcast}. The negative impact of any of the aforementioned attacks is proportional to the number of compromised devices. It is worth noting that, in IoT/CPS systems, adversaries can compromise, control, and manipulate physical equipment, which may lead to physical consequences such as equipment sabotage, power outage, vehicles collisions, or workers injury~\cite{CPS_Sec_Survey}.  The aforementioned security risks call for resilient, robust, and ubiquitous security countermeasures to safeguard IoT/CPS networks from large-scale malware attacks.

\section{Prior Art \& Contributions}

One major research direction is to develop lightweight security countermeasures for IoT/CPS devices. Per-device IoT/CPS security can be implemented either in hardware~\cite{hasan2019protecting} or in software~\cite{darabian2020opcode, nguyen2020psi}. However, many IoT/CPS devices are too constrained (e.g., storage, energy, and computational power) to implement such per-device countermeasures. Furthermore, due to the massive number of devices, implementation of hardware solutions and licensing of software countermeasures may impose overwhelming monetary costs. Hence, it is infeasible to ubiquitously safeguard $100\%$ of the devices against malware intrusion/infection~\cite{Junaid, Epidemics1, Epidemics2}. Articulated differently, interim infection of some devices is inevitable in large-scale massive IoT/CPS systems. Hence, timely detection and treatment of malware is the security challenge in large-scale IoT/CPS networks such that malware outbreak is prevented. Otherwise the malware infection goes out of control and large populations of devices are compromised. 

To detect compromised devices, the authors in \cite{seda, yan2019eapa} propose software attestation schemes to insure the integrity of the running software and configuration of IoT devices. However, the attestation schemes in \cite{seda, yan2019eapa} are centralized, which may impose overwhelming overhead traffic and delay to detect compromised devices. The work in \cite{Epidemics1} proposes a game theoretic approach to select the devices that install anti-malware programs such that an epidemic outbreak is prevented. However, the proposed mechanism in \cite{Epidemics1} is based on a fully mixed epidemic model,\footnote{A fully mixed epidemic model assumes that an infection (e.g., malware) can be directly transmitted from any node in the network to any other node in the network.} which is not adequate for wireless IoT networks. Accounting for the physical layer parameters of wireless networks, the authors in~\cite{Junaid} propose periodic software patching for IoT/CPS devices to eliminate potential malicious codes to combat botnet formation. However, the technique proposed in~\cite{Junaid} is oblivious to the device status, which may lead to unnecessary disruption for the IoT/CPS operation as a price for patching healthy devices. Furthermore, compromising a device shortly after being attested and/or patched may grant adversaries enough time to launch malicious attacks. Such scheduled software attestation/patching problems are more acute when employing wireless technologies such as NOMA and URLLC due to the accelerated epidemic infection rate.

\begin{table*}[t]\caption{Table of Notations}
	\centering
	\begin{center}
		\resizebox{0.7 \textwidth}{!}{
			\renewcommand{\arraystretch}{1.4}
			\begin{tabular}{ {c} | {c} }
				\hline
				\hline
				\textbf{Notation} & \textbf{Description} \\ \hline
				$\Phi$; $\lambda_r$; $r_r$ & the set of IoT/CPS devices locations; their intensity; their communication range \\ \hline
				$\Psi$; $\lambda_f$; $r_f$ & the set of firewalls locations; their intensity; their communication range \\ \hline
				$\Xi=\Phi\setminus\Theta$ & the set of susceptible devices locations \\ \hline
				$\Theta=\Phi\setminus\Xi$ & the set of protected devices locations \\ \hline
				$G=(\Phi,E)$ &  the RGG representation of IoT/CPS network with vertex set $\Phi$ and edge set $E$ \\ \hline
				$\mathcal{I}=\{\Xi ,\mathcal{E}\}$ & the infection susceptible graph with vertex set $\Xi$ and edge set $\mathcal{E}$\\  \hline
				$\theta_{G}(\lambda_r,r_r)$ & probability of percolation in $G$\\ \hline
				$\theta_{\mathcal{I}}(\lambda_f,r_f,\lambda_r,r_r)$ & probability of percolation in $\mathcal{I}$\\ \hline
				$\lambda_f^c$ & critical density of firewalls\\ \hline
				$\mathcal{L}_h$; $\mathcal{L}_s$; $\mathcal{L}_s^d$ & hexagonal lattice; square lattice; dual of square lattice\\ \hline
				$K$, $K_{\mathcal{L}_h}$, $K_{\mathcal{L}_s}$, $K_{\mathcal{L}_s^d}$, $K_{\mathcal{I}}$  & connected component in $G$, $\mathcal{L}_h$, $\mathcal{L}_s$, $\mathcal{L}_s^d$ and  $\mathcal{I}$, respectively \\ \hline
				$\delta_{\text{sec}}$;  $\delta_{\text{sec}}^c$ & the percentage of protected devices; critical percentage of protected devices \\  \hline\hline
		\end{tabular}}
	\end{center}
	\label{tab:TableOfNotations}
\end{table*}



{To overcome the aforementioned problems, the authors in~\cite{elsawy_mag_2020} propose a novel countermeasure denoted as ``\textbf{\em spatial firewalls}''. The spatial firewalls are wireless devices, with sufficient computational power, energy resources, and memory, to store, execute, and frequently update anti-malware and intrusion detection programs. Spatial firewalls can be edge servers, access points, or capable IoT/CPS devices, which are randomly deployed in the network to analyze the wireless traffic in order to detect and thwart emerging malware infections. However, the exposition in  \cite{elsawy_mag_2020} is based on simulations, which lacks the mathematical details that are necessary to prove the concept, assess, and design spatial firewalls. In this context, we develop a rigorous mathematical framework to assess and design spatial firewalls. In order to account for the underlying limited-range wireless connectivity for the firewalls and IoT/CPS devices, we utilize percolation theory on random geometric graphs (RGG) for the developed mathematical framework. }

It is worth noting that percolation theory on RGG is widely used to assess information dissemination and global network connectivity in wireless sensors networks~\cite{GilbertDisk1, GilbertDisk2},  robot swarms~\cite{GilbertDisk8}, high altitude platforms~\cite{GilbertDisk10}, and connected unmanned aerial vehicles~\cite{GilbertDisk9}. Percolation models are also used to study information dissemination in cognitive networks~\cite{Swami, Yemini}.
Note that the models in \cite{GilbertDisk1, GilbertDisk2,GilbertDisk8,GilbertDisk9,GilbertDisk10, Swami, Yemini} assume proximity based wireless communications, which does not account for the aggregated network interference. Interference-aware percolation models are developed in~\cite{SINR2, SINR3}, where nodes communicate if and only if the signal-to-interference-plus-noise ratio exceeds a certain threshold. Percolation theory is also used in~\cite{Pinto1, Rahul1} to study private information dissemination in the presence of eavesdroppers. 

{In this paper, we utilize percolation theory on RGG to characterize wireless malware propagation in IoT/CPS networks, and hence,  prove the concept and assess the spatial firewalls solution. That is, we mathematically prove the existence of a critical density of firewalls beyond which malware outbreak becomes impossible. To this end, we find a tractable upper-bound on the critical density of firewalls that is required to safeguard large-scale IoT/CPS networks against malware epidemics. In addition, we present several insights for the design of spatial firewalls. The contributions of this paper can be summarized in the following points 
\begin{itemize}
	\item We define the infection susceptible graph (ISG) to characterize the risk malware propagation in large-scale IoT/CPS networks.
		\item Using percolation theory along with the ISG, we prove that properly designed firewalls are capable to safeguard large-scale IoT/CPS networks from malware outbreak irrespective of the infection propagation rate. 
	\item We derive a tractable upperbound for the critical density of spatial firewalls that is required to safeguard large-scale IoT/CPS networks against malware epidemics.  
		\item We analytically characterize the IoT/CPS communications range that allows network connectivity while prohibiting malware epidemics.
		\item  We provide several insights on the percentage of IoT/CPS devices that are protected via the spatial firewalls.     
\end{itemize}}
It is worth noting that the spatial firewalls represents one layer of the IoT cybersecurity countermeasures. In the IoT era, specially when the devices are simple, ultra-dense, and managed by general public, the security problem is not a single-sided IT problem. Instead, cybersecurity in IoT is a multi-dimensional problem that should be collaboratively solved by wireless communication experts, machine learning experts, hardware designers, software developers, IT experts, in addition to IoT consumers by raising awareness regarding cybersecurity threats. This paper focuses on the wireless communications and networking aspect of the IoT and provides a proactive countermeasure that eliminates the risk of large-scale diffusion of malware epidemics.
\subsection{Paper organization}
The rest of the paper is organized as follows. Section \ref{sec:SysMod} presents the system model and formulates the large-scale malware epidemic problem in terms of graph and percolation theory. Section \ref{Phase_transition} proves the concept of spatial firewalls and shows the existence of a critical density for the spatial firewalls that safeguards large-scale IoT/CPS from malware epidemics. Section \ref{Sec:upper} discusses different design schemes for the spatial firewalls.  Simulation and numerical results are presented in Sec. \ref{simulations}. Finally, concluding remarks are given in Sec. \ref{conclusion}. {For the ease of mathematical exposition, frequently used symbols are summarized in Table~\ref{tab:TableOfNotations}.}

\section{System Model} \label{sec:SysMod}

We consider large-scale IoT/CPS network with ad hoc topology. In particular, the IoT/CPS devices are assumed to be scattered in $\mathbb{R}^2$ according to a homogeneous Poisson point process (PPP) $\Phi\equiv\{x_0, x_1, \cdots, x_k, \cdots\} \subset\mathbb{R}^2$ with intensity $\lambda_r$. We model the locations of the IoT devices as a PPP  with density $\lambda_r$. Note that the PPP is commonly utilized and widely accepted in the literature to model wireless networks due to its tractability and practical significance \cite{Junaid,PPP1,PPP2,PPP3,PPP4,PPP5}. The IoT/CPS devices can establish bidirectional device-to-device (D2D) links if they are within the wireless communication ranges of each other. It is assumed that all devices have the same wireless communication range of $r_r$ meters.  All IoT/CPS devices are assumed to be too constrained to install and execute anti-malware programs. Hence, the D2D links can be used for legitimate code dissemination or exploited for malware infection propagation. We assume autonomous malware worms in which compromised devices are infection threats for all of their connected neighbors~\cite{Nature_worm,wireless_worm }. Exploiting multi-hop D2D connectivity, the malware infection may diffuse to large population of devices and create an epidemic outbreak. 

\begin{figure*}
	\centering
	\includegraphics[width=0.99\textwidth,  trim={0cm 2.5cm 0cm 2.2cm},clip]{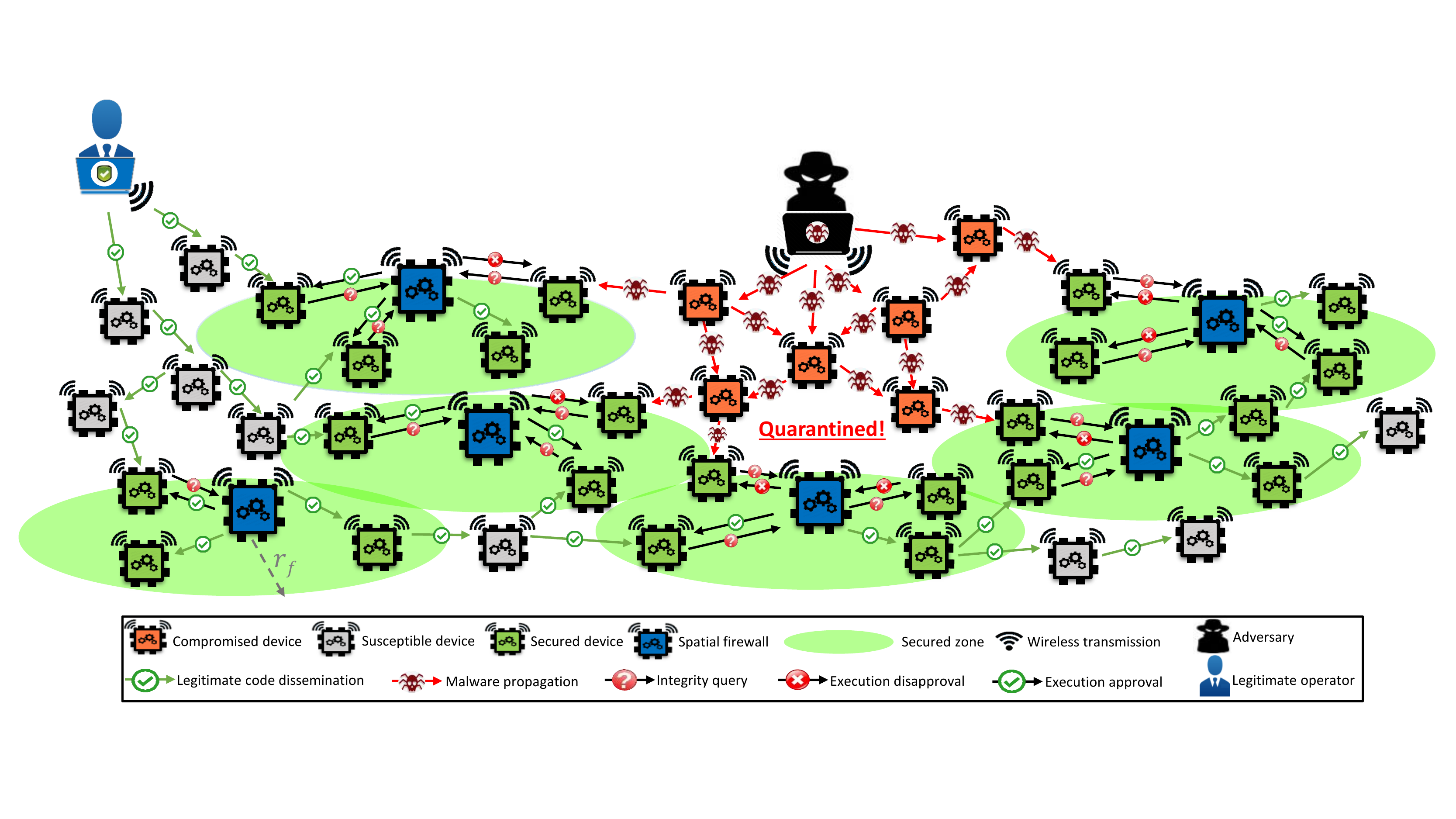}
	\caption{Illustration of spatial firewalls operation.}
	\label{fig:SPFW}
\end{figure*}

To secure such large-scale IoT/CPS network from malware epidemic outbreak, spatial firewalls are randomly deployed according to an independent PPP $\Psi\equiv\{a_0, a_1, \cdots, a_k, \cdots\}\subset \mathbb{R}^2$ with intensity $\lambda_f$. Firewalls are computationally capable devices (e.g., edge computing device, access point, high-end IoT/CPS devices) that are equipped with the state-of-the-art anti-malware programs~\cite{malware_detection1,malware_detection2,malware_detection3}.  Each firewall is assumed to have a communication range of $r_f$ meters, and hence, each firewall creates a {\em secured zone} of radius $r_f$ around itself. Due to the higher transmit power and better signal capture capabilities of firewalls, it is assumed that the firewalls have larger communication and detection ranges than the IoT/CPS devices (i.e., $r_f \geq r_r$). IoT/CPS devices within secured zones inquire the firewall about software codes received from the wireless interface. The firewalls scan inquired codes for security threats. If the code is legitimate and free from malware threats, the firewalls approves it. Otherwise, the code is disapproved and reported to the security administration for further action.\footnote{For instance, the security administration may need to launch a patching campaign to recover compromised devices and prevent further propagation of the malware to other vulnerable devices.}
Consequently, IoT/CPS devices within secured zones are protected from malware infection and do not participate in malware propagation. On the other hand, devices outside secured zones have no direct connectivity with firewalls, and hence, they opt to directly execute and relay the codes received from the wireless interface. Hence, devices that are outside secured zones are susceptible to malware infection and may participate to malware propagation. In practice, we can expect that spatial firewalls are continuously updated, monitored and maintained with highly skilled personnel. Hence, in analysis, we assume that spatial firewalls are never compromised with attackers and can ensure almost $100\%$ security. The operation of the spatial firewalls in large-scale IoT/CPS is depicted in Fig.~\ref{fig:SPFW}. As shown in the figure, if the firewalls are dense enough, the collective impact of secured zones can thwart malware outbreak by spatially quarantining malware infections within finite region. The firewalls can then report to the security administration about detected malware for localized patching and treatment of compromised devices. As shown in Fig.~\ref{fig:SPFW}, the IoT/CPS network is composed from three types of devices:
\begin{itemize}
	\item	\textit{\textbf{Spatial firewalls}} are capable devices, equipped with the state-of-the-art anti-malware programs, that are spatially distributed across the network.
	\item	\textit{\textbf{Protected devices}} are IoT/CPS devices that fall within the secured zone of a firewall, and hence, can not be infected with malware and do not participate in malware propagation. 
	\item \textit{\textbf{Susceptible devices}} are IoT/CPS devices that fall\ outside the secured zone of a firewall, and hence, can be compromised and may participate in malware propagation. 
\end{itemize}

%

\subsection{Mapping to Graph and Percolation Theory}

To study and characterize malware propagation in IoT/CPS networks and assess the impact of spatial firewalls, we utilize graph and percolation theory. In particular, the IoT/CPS network is mapped to a RGG, denoted as $G=\{\Phi,E\}$, where the devices $\Phi$ are mapped to the graph vertices. Accounting for the limited wireless D2D communications range of $r_r$, the set of edges $E$ is defined as
\begin{equation}\label{eq:edge}
E=\left\{\overline{x_ix_j}: \left\|x_i-x_j\right\|\leq r_r, \; x_i,x_j \in \Phi \right\},
\end{equation}
where $\left\|\cdot\right\|$ denotes the Euclidean norm and $\overline{x_ix_j}$ is the edge connecting $x_i$ and $x_j$. The edges defined in \eqref{eq:edge} represent bidirectional direct (i.e., one hop) D2D connectivity between devices. The bidirectional links in $E$ can be used for legitimate traffic dissemination or malware propagation. An infection from a device in $G$ can reach its direct D2D neighbor devices in one hop. Furthermore, an infection from a device in $G$ can also reach non-neighbor distant devices through multi-hop connectivity if there exists a route in $E$ that connects the compromised device to the distant device. However, due to the random devices locations and limited wireless D2D range, a compromised device does not imply an infection threat to all other devices in $G$. This is because there might not be a multi-hop route in $E$ that connects the compromised device to all other devices in $G$. The mutual infection threat between devices in $G$ is specified through the connected components, which are defined as: 
\begin{definition}[Connected Component] 
\label{def:1} A connected component is a sub-graph $K \subseteq G(\Phi,E)$ with the largest possible devices such that, within $K$, any device $x_i \in K$ can always find a multi-hop route through a set of consecutive edges in $E$ to any other device $x_j\in K$, $i\neq j$. Consequently, a malware infiltration to any device $x_i\in K$ represents an infection threat for all devices $x_j\in K$ for $i\neq j$. \\
\end{definition}
Based on Definition~\ref{def:1} and the ability of malware infection to exploit multi-hop connectivity, a compromised device is an infection threat to all devices within its own connected component. Hence, the infection threat is directly proportional to the size of connected components. Small values of $\lambda_r$ and/or $r_r$ lead to sparse vertices in $\Phi$, and hence, the graph $G$ will be consisting from several disjoint small connected components. In such case, there is no risk of an epidemic outbreak due to the lack of multi-hop wireless connectivity that connects large population of devices. Increasing $\lambda_r$ and/or $r_r$, the connected components start to merge together into larger components and an infection becomes threatening to a larger number of devices. Sufficiently high  $\lambda_r$ and/or $r_r$ create a \textit{giant component} that connects infinite number of devices \cite{Haengi_book, Haengi2}. The existence of a giant component implies the risk of an epidemic outbreak that gets out of control and compromise large-population of IoT/CPS devices. Characterizing the network parameters that lead to the existence/absence of the giant component is the core focus of \textit{percolation} theory. Formally, the percolation probability on the graph $G$, as a function of $\lambda_r$ and/or $r_r$, is defined as 

\begin{definition}[Percolation Probability] 
\label{def:2} 
Percolation probability defines the probability of existence of infinitely large connected component $K \subseteq G$, defined as
\begin{align}\label{Percolation prob}
\theta_{G}(\lambda_r,r_r)=\mathbb{P}\{|K|=\infty\}, 
\end{align} 
where $|\cdot|$ denotes the set cardinality. A non-zero percolation probability $\theta_{G}(\lambda_r,r_r)>0$ defines the super-critical regime in which the network percolates and a giant component exists.  On the other hand, a zero percolation probability  $\theta_{G}(\lambda_r,r_r)=0$ defines the sub-critical regime with no percolation and no giant components.
\end{definition}
In the context of the malware infection in IoT/CPS network, the super-critical regime implies the risk of having a large connected population of devices that are vulnerable to an epidemic outbreak if a single device is compromised. The relative values of $\lambda_r$ and $r_r$ that lead to super-critical regime operation and raise the risk of malware epidemic in the depicted IoT/CPS network is defined in the following proposition. 

\begin{lemma} \label{dd}
The IoT/CPS network operates in the super-critical regime $\theta_G(\lambda_r,r_r)>0$, and hence, is susceptible to malware epidemic if and only if 
	\begin{equation}
	\label{eq:connectivity1}
	\lambda_{r} \geq \frac{\lambda_{c}(1)}{r_r^2},
	\end{equation}
	where $\lambda_{c}(1)\approx 1.44$. 
\begin{IEEEproof}
The proof is similar to \cite[Chapter 2]{Roy} which characterizes continuum percolation on PPP networks with homogeneous communication ranges. 
\end{IEEEproof}
	\end{lemma}
	
	If (\ref{eq:connectivity1}) is not satisfied, the IoT/CPS network is physically immune to malware epidemics. That is because if the defence mechanism of an IoT device is beaten by the attacker, the largest infected region will still be finite.
\begin{remark}
	\label{remark22}
$\lambda_{c}(1)$ defined in Lemma~\ref{dd} is the critical (i.e., minimum) intensity of nodes required for continuum percolation in homogeneous PPP network with a normalized communication range 1. There is no known exact value for $\lambda_{c}(1)$ in the literature. However, there exists some useful approximations in literature such as  $\lambda_{c}(1)\approx 1.44$ \cite{Quintanilla_2000}. There are also analytically derived lower and upper bounds: $0.768<\lambda_c(1)<3.37$ \cite{Roy,Yeh}.
\end{remark}

If the condition defined in \eqref{eq:connectivity1} of Lemma~\ref{dd} is not satisfied, then the IoT/CPS network is physically immune to malware epidemics due to the lack of multi-hop D2D connectivity that can be exploited to transfer malware infection to large-population of devices. Otherwise, the IoT/CPS network is at risk of malware epidemic and the spatial firewalls countermeasure is required. Note that in dense IoT/CPS networks, the condition in \eqref{eq:connectivity1} is usually satisfied. 

\begin{figure*} [t!]
	\begin{subfigure}{.5\textwidth}
		\centering
		\includegraphics[width=  3.0 in, trim={3.9cm 1cm 2.5cm 0.8cm},clip]{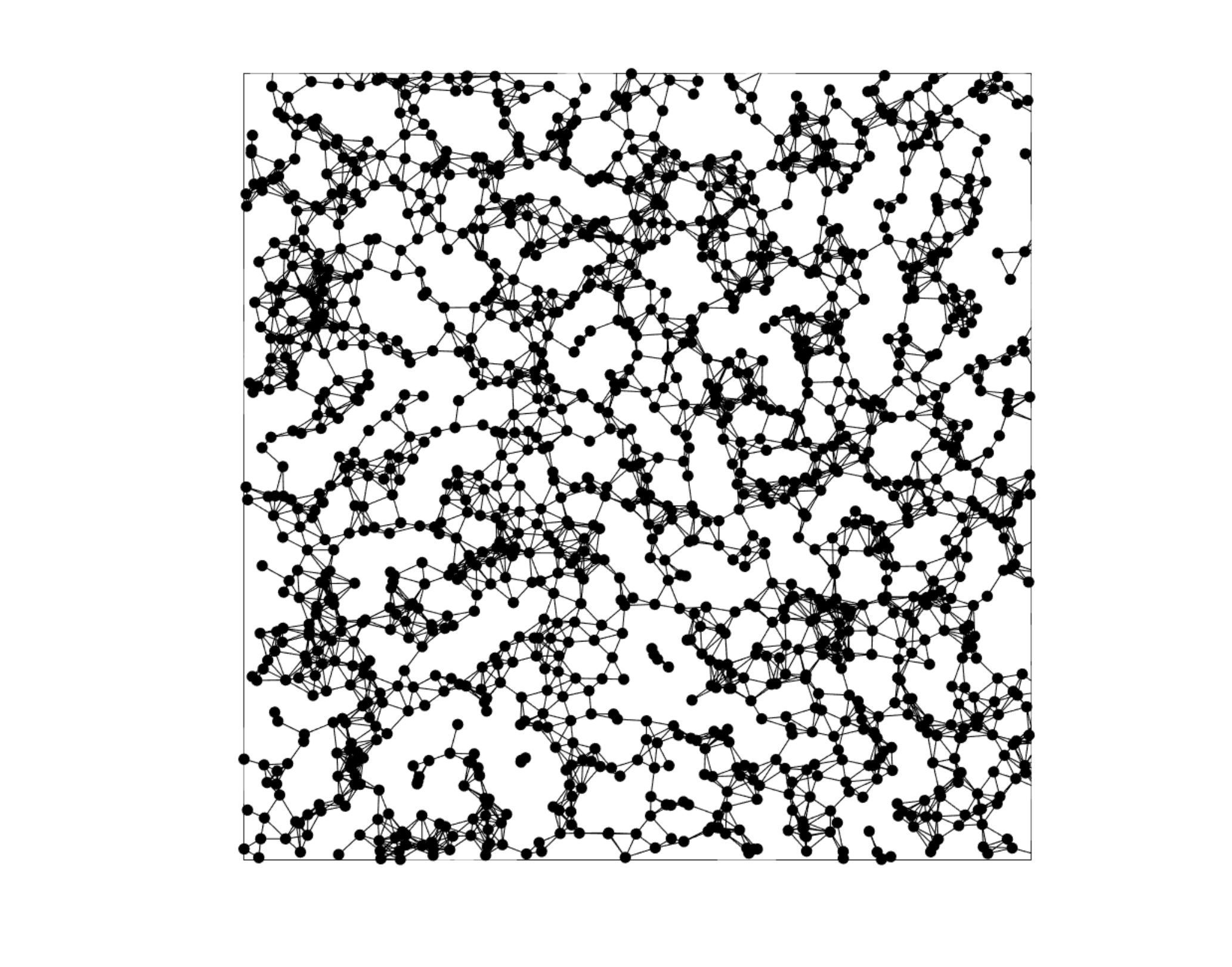}
		\caption{}
		\label{fig3:a}
	\end{subfigure}
	\begin{subfigure}{.5\textwidth}
		\centering
		\includegraphics[width=  3.0 in, trim={3.9cm 1cm 2.5cm 0.8cm},clip]{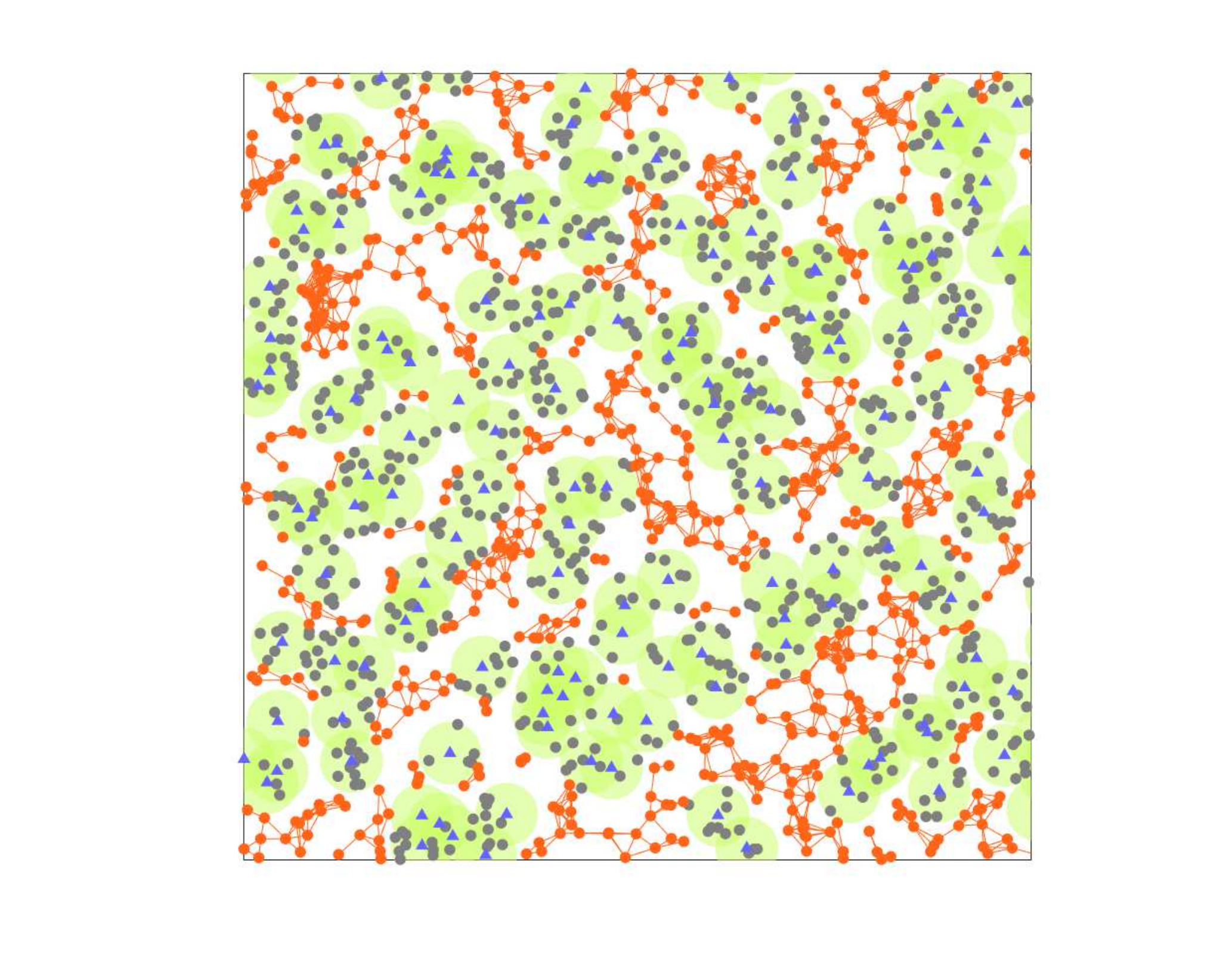}
		\caption{}
		\label{fig3:d}
	\end{subfigure}
	\caption{The left figure shows a realization of $G(\Phi,E)$ before the deployment of spatial firewalls. The right figure shows the impact of spatial firewalls (blue triangles with green secured zones), which splits $G(\Phi,E)$ into the ISG $\mathcal{I}=\{\Xi ,\mathcal{E}\}$ (orange connected nodes) and protected deceives $\Theta$ (grey nodes).}
	\label{fig:System_model_big}%
\end{figure*}

As discussed earlier, the spatial firewalls introduce spatial secured zones that protect some IoT/CPS devices and thwart malware propagation.\footnote{Different form traditional reactive cybersecurity countermeasures, the proposed spatial firewall solution is a proactive networking solution that eliminates network-wide malware infection risks rather than reacting to local attacks. To defend against local attacks (i.e., within single or multiple proximate susceptible devices), there should be complementing reactive security countermeasure.} 
 To incorporate the impact of spatial firewalls to the mathematical framework, the vertices $\Phi$ in the RGG $G$ are further divided into susceptible devices $\Xi \subseteq \Phi$ and protected devices $\Theta \subseteq \Phi$ such that $\Xi  \cup \Theta= \Phi$ and $\Xi  \cap \Theta =\emptyset$. The protected set $\Theta =\{x_i \in \Phi:  \min\limits_{a_j\in \Psi}\left\|x_i-a_j\right\|\leq r_f \}$, where $\min\limits_{a_j\in \Psi}\left\|x_i-a_j\right\|$ is the minimum distance between $x_i$ and all firewalls in $\Psi$. Hence, $\Theta$ contains all the devices that are located within the secured zones of the firewalls, i.e., the devices that can neither be infected nor participate in malware propagation. On the other hand, the susceptible set $\Xi  =\{x_i \in \Phi:  \min\limits_{a_j\in \Psi}\left\|x_i-a_j\right\| > r_f\}$ contains the devices that are located outside the secured zones of all firewalls. {A pictorial illustration for a realization of $G=(\Phi,E)$ before and after deploying spatial firewalls is shown in Fig.~\ref{fig:System_model_big}.}

To characterize the impact of spatial firewalls, we define the {\em infection susceptible graph} (ISG) $\mathcal{I}=\{\Xi ,\mathcal{E}\}$, with all susceptible devices in $\Xi$ and set of edges $\mathcal{E}$, which contains all the D2D links that can be exploited for malware infection propagation. The set $\mathcal{E}$ is defined as
\begin{align}
\mathcal{E}&\overset{(a)}{=}\left\{\overline{x_ix_j}: \left\|x_i-x_j\right\|\leq r_r, \; x_i, x_j \in \Xi \right\}.
\label{e11}
\end{align}
The ISG is illustrated in Fig.~\ref{fig3:d}, which highlights all susceptible deceives and their D2D connectivity. It is worth noting that an infection cannot propagate from a device in $\Xi$ to a device in $\Theta$, or vice versa, due to the firewall protection for all devices in $\Theta$. Hence, the definition in \eqref{e11} for the edges in $\mathcal{E}$ is restricted to the susceptible devices in $\Xi$.

 It is clear that the ISG is a subset of the IoT/CPS network graph  $\mathcal{I}\subseteq G$. At the absence of spatial firewalls (i.e., $\lambda_f=0$), all the devices are susceptible to infection, and hence, the ISG coincides with the IoT/CPS network graph $\mathcal{I} = G(\Phi,E)$. {Deploying spatial firewalls splits the set $\Phi$ into protected $\Theta$ and susceptible $\Xi$ devices.} The ISG $\mathcal{I}$ can be constructed by removing the vertices in $\Theta$ and their associated edges from $G(\Phi,E)$. Given that $G$ operates in the super-critical regime, the objective is to deploy sufficiently dense spatial firewalls (i.e., $\lambda_f$) such that the ISG  $\mathcal{I}\subseteq G$ operates in the sub-critical regime. Note that the sub-critical regime operation of $\mathcal{I}\subseteq G$ implies that the risk of malware epidemic is eliminated. Consequently, the IoT/CPS network is safeguarded from malware epidemics regardless of the malware infection rate. Let $K_\mathcal{I}\subseteq  \mathcal{I} $ be the largest connected component in the ISG $\mathcal{I}$ and let $ \theta_{\mathcal{I}}(\lambda_f,r_f,\lambda_r,r_r)  =\mathbb{P}\{|K_\mathcal{I}|=\infty\}$ be the percolation probability of the ISG $\mathcal{I}$. Then, the design objective of the spatial firewalls is formally defined as 

 \begin{equation}
 \begin{aligned}
 & \text{minimize}
 & & \lambda_f \\
 & \text{subject to}
 & & \theta_{\mathcal{I}}(\lambda_f,r_f,\lambda_r,r_r) = 0,
 \end{aligned}
 \label{object}
 \end{equation}

 To minimize the monetary cost (e.g., deployment and/or anti-malware licensing) of spatial firewalls, it is desirable to find the minimum intensity of firewalls that safeguards the IoT/CPS network against malware epidemics. In the notion of percolation theory, the optimal $\lambda^*_f$ is denoted as the critical density for percolation. 


\section{Proof of Concept}
\label{Phase_transition}


 This section proves the concept of spatial firewalls by showing that there exists a phase transition for the percolation probability of the ISG. In particular, there is a critical intensity of spatial firewall $\lambda_f^c$  below which the ISG operates in the super-critical regime. Hence, if the firewalls are not dense enough, the percolation probability is non-zero and the risk of malware epidemic exists. If the intensity of firewalls is above the critical intensity $\lambda_f^c$, the ISG operates in the sub-critical regime, which eliminates the risk of malware epidemic by enforcing a zero percolation probability. To complete the proof of concept, we show that the percolation probability is monotonically decreasing in $\lambda_f$, which proves the phase transition at a critical intensity $\lambda_f^c$. Hence, the critical intensity $\lambda_f^c$ is the optimal density that minimizes \eqref{object}.
 
 For the sake of organized presentation, Section~\ref{sec:sub} presents the sub-critical regime operation for the ISG, which proves the effectiveness of the spatial firewalls. Then,  Section~\ref{sec:super} presents the super-critical regime operation for the ISG, which proves the need for dense enough spatial firewalls. Last but not least, Section~\ref{sec:mon} completes the proof of concept by showing the monotonicity of the percolation probability in $\lambda_f$.

 \subsection{Sub-critical regime} \label{sec:sub}
 
\begin{figure*} [t!]
	\begin{subfigure}{.5\textwidth}
		\centering
		\includegraphics[scale=0.5]{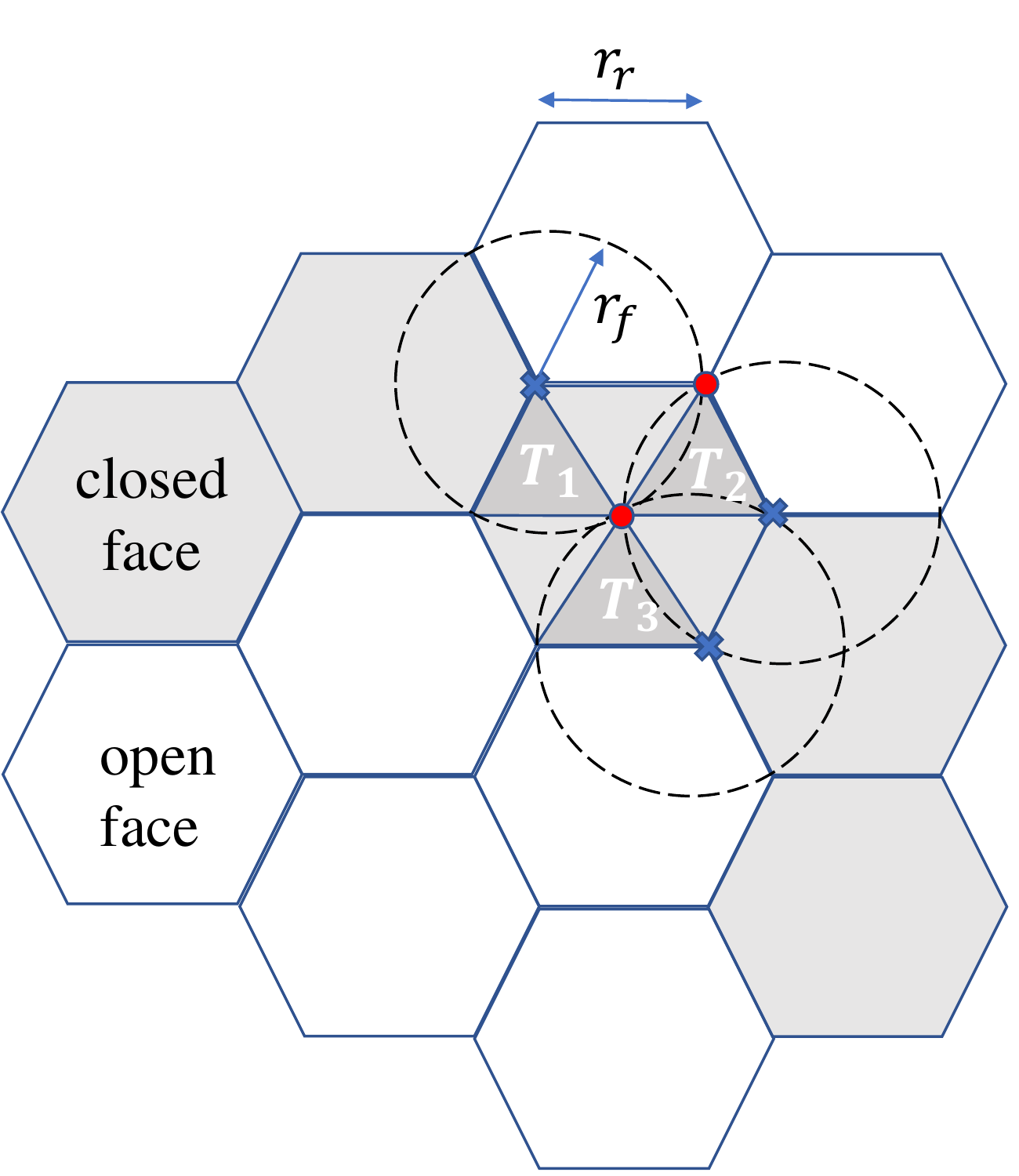}
		\caption{A closed face illustration.}
		\label{Fig:closed face}
	\end{subfigure}
	\begin{subfigure}{.5\textwidth}
		\centering
		\includegraphics[scale=0.15]{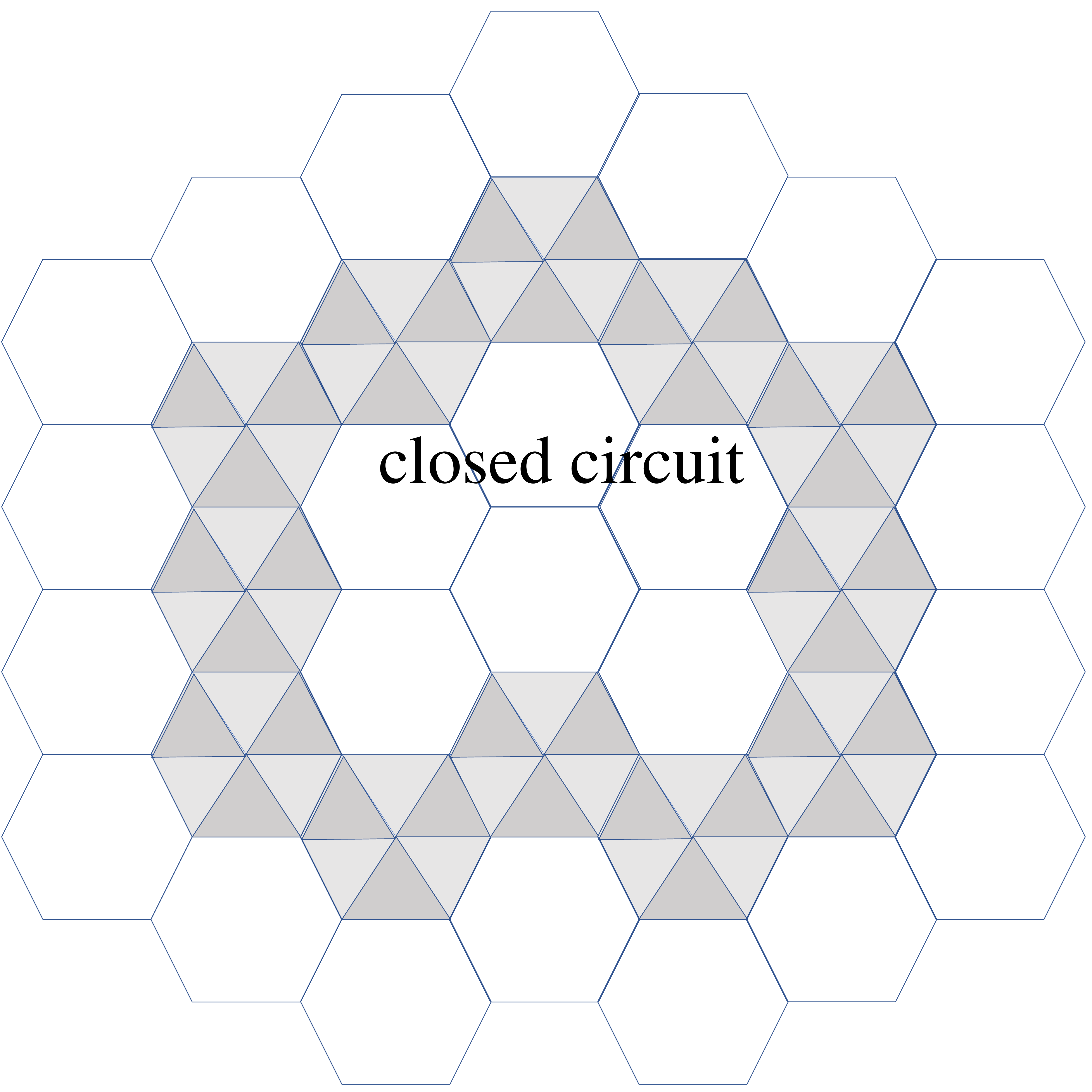}
		\caption{A closed circuit around the origin.}
		\label{Fig 3}
	\end{subfigure}
	\caption{Illustrating the concept of closed face and closed circuit in  the hexagonal lattice $\mathcal{L}_h.$}
	\label{Fig:sub}
\end{figure*}

 This section proves that sufficiently dense firewalls enforce a sub-critical regime operation for the ISG, which safeguards the IoT/CPS networks against malware epidemics.  To find sufficient conditions for spatial firewalls intensity that enforces sub-critical regime operation for the ISG, a worst case scenario of $r_f=r_r$ is assumed. Such sufficient firewalls intensity would also enforce sub-critical regime for the general case of $r_f\geq r_r$. For tractable analysis, the common practice in percolation theory is to study continuum percolation in RGG by mapping them to discrete lattices. Inspired by \cite{Pinto1}, we prove the sub-critical regime operation by mapping the ISG to hexagonal lattice as defined in the sequel. \\
\textbf{Mapping to a Hexagonal Lattice:} Let $\mathcal{L}_h$ be a hexagonal lattice with a side equal to the D2D communication range $r_r$, which is also equal to the secured zone radius is (i.e., $r_f=r_r$).  Let $\mathcal{H}$ denote a randomly selected hexagon, also denoted as a face, in $\mathcal{L}_h$. Depending on the firewalls occupancy, a face $\mathcal{H}$ can be either open or closed, as explained next. 
\begin{definition}[Closed/Open face in $\mathcal{L}_h$]
	\label{def_face}
	Let $\left\{T_i\right\}^3_{i=1}$ denote three non-adjacent equilateral triangles within a face $\mathcal{H}$ as shown in Fig.\ref{Fig:closed face}. Then, the face $\mathcal{H}$ is said to be closed if each of these triangles is occupied with at least one firewall. Otherwise, the face $\mathcal{H}$ is denoted as \textit{an open face}.
\end{definition}
{Definition \ref{def_face} is chosen such that the absence of open
face percolation in $\mathcal{L}_h$ assures no continuum percolation in the ISG $\mathcal{I}=(\Xi,\mathcal{E})$. In particular, open face percolation is obstructed by closed faces. In our setup, a closed face defines a protected geographical region (i.e., by secured zones of firewalls) that cannot be crossed by a malware infection. More precisely, due to the union of the secured zones of the firewalls within the triangles $\left\{T_i\right\}^3_{i=1}$, an infected device within the vicinity of a closed face will not have any susceptible device within its D2D reach through the closed face.} Articulated differently, there could not be susceptible edges in $\mathcal{E}$ of the ISG that passes through a closed face in $\mathcal{L}_h$. A sequence of connected closed faces form \textit{a closed path}, which further extends the firewalls spatial protection to larger connected (i.e, no gaps for malware propagation) geographical region. A path that starts and ends at the same face is denoted as a \textit{closed circuit}. { A closed circuit on $\mathcal{L}_h$ implies no open face percolation on $\mathcal{L}_h$, which also implies finite connected component in $\mathcal{I}=(\Xi,\mathcal{E})$. Hence, closed circuit on $\mathcal{L}_h$  means spatially quarantined (i.e., surrounded) malware. A pictorial illustrations of a closed face and a closed circuit are shown in Fig. \ref{Fig 3}. }

As illustrated above, malware infection is obstructed by closed faces. Hence, an infection that originated within a closed circuit is spatially quarantined within the connected component of the infected device. 
Due to the stationarity of the PPP, there is no loss of generality to assume that the infection originates at the device located at the origin. Hence, it is sufficient to prove that the origin is surrounded by a closed circuit to prove that the ISG operates in the sub-critical regime. The coupling between the hexagonal lattice $\mathcal{L}_h$ and the ISG $\mathcal{I}$ is formally stated and proved in the following lemma.

\begin{lemma} [Hexagonal lattice coupling]
	\label{lemma1}
	Let $K_\mathcal{I}(0)\subseteq \mathcal{I}(\Xi, \mathcal{E})$ and $K_{\mathcal{L}_h}(0)\subseteq \mathcal{L}_h$ denote connected components around the origin in, respectively, the ISG $\mathcal{I}$ and the hexagonal lattice $\mathcal{L}_h$. If $K_{\mathcal{L}_h}(0)$ is surrounded with closed-circuit $\mathcal{C}(0)$ in $\mathcal{L}_h$, then  $K_\mathcal{I}(0)$ is finite.
	\begin{IEEEproof} A closed circuit around the origin implies a finite number of open faces on the inner side of the circuit. Consequently, $|K_{\mathcal{L}_h}(0)|<\infty$ and the region covered by $K_{\mathcal{L}_h}(0)$ involves a finite number of vertices of $\mathcal{I}(\Xi,\mathcal{E})$. Hence, to prove that $K_\mathcal{I}(0)$ is finite, it is sufficient to prove that no edge of $\mathcal{I}(\Xi,\mathcal{E})$ crosses $\mathcal{C}(0)$. Let us consider an extreme scenario for the closed face with the worst spatial setup for the three firewalls and IoT/CPS devices, shown in Fig.\ref{Fig:closed face}. In particular, assume each of the triangles $\left\{T_i\right\}^3_{i=1}$ has only one firewall in the shown worst case locations such that their secured zones provides minimum protection (i.e., coverage) of the closed face. Furthermore, let us consider the most advantageous location for the IoT/CPS devices for malware infection propagation as shown Fig.\ref{Fig:closed face}. Recall that the side of each equilateral triangle is $r_r$, then we have one of the following two scenarios. 1) If the two devices are within the D2D communication range of each other, then one of them should be in $\Theta$ (i.e., within the secured zone of one or more of the three spatial firewalls). 2) If both devices are in $\Xi$ (i.e., both of them are out of the range of the three firewalls), then they are out of the D2D communication range of each other (i.e., the condition in \eqref{e11} is not satisfied). The example shown in Fig.\ref{Fig:closed face} shows that even in the worst case spatial setup of firewalls, an infection cannot bypass the closed face.	 Therefore, no edge in $\mathcal{E}$ can cross $\mathcal{C}(0)$, and hence we conclude that a finite $|K_{\mathcal{L}_h}(0)|$ leads to a finite $|K_\mathcal{I}(0)|$. 
		
	\end{IEEEproof}
\end{lemma}

Exploiting the mapping to the hexagonal lattice and the coupling introduced in Lemma~\ref{lemma1}, we can state the main result of this section in the following  proposition 

\begin{prop}[Sufficient condition for zero percolation on ISG]
	\label{Prop1}
	For  given $\lambda_r>0$ and $r_r>0$, the ISG operates in the sub-critical regime (i.e., $\theta_\mathcal{I}(\lambda_f,r_f,\lambda_r,r_r)=0$)  if
	\begin{equation}
	\label{eq4}
	\lambda_f>\frac{3.65}{r_r^2}.
	\end{equation}

\begin{IEEEproof} 
Referring to the results in Lemma~\ref{lemma1}, a closed circuit $\mathcal{C}(0)$ on the hexagonal lattice $\mathcal{L}_h$  implies sub-critical regime operation of the ISG $\mathcal{I}$. Based on the results in  \cite{Bollobas2006},  the origin is almost surely (a.s.) surrounded with closed circuit $\mathcal{C}(0)$ in $\mathcal{L}_h$ if
\begin{equation}
\label{eq6}
P(\mathcal{H}\text{ is closed})>\frac{1}{2}.
\end{equation}
From the PPP properties of the spatial firewalls we have 
\begin{align}
P(\mathcal{H} \text { is closed }) &=\mathrm{P}\left(\bigcap_{i=1,2,3}\left|T_{i} \cap \Psi\right| \geq 1\right) \notag \\
&=\left(1-\mathrm{P}\left(\left|T_{1} \cap \Psi\right|=0\right)\right)^{3} \notag \\
&=\left(1-e^{-\lambda_{f} \frac{\sqrt{3}}{4} r_{r}^{2}}\right)^{3}.
\label{eq7}
\end{align}
Substituting \eqref{eq7} back in \eqref{eq6}, we conclude that the origin is a.s surrounded with closed circuit $\mathcal{C}(0)$ in $\mathcal{L}_h$ if
	\begin{equation}
\label{eq5}
\left(1-e^{-\lambda_{f} \frac{\sqrt{3}}{4} r_r^{2}}\right)^{3} >\frac{1}{2}.\end{equation}
Rearranging the terms of (\ref{eq5}) and following to the statement of Lemma \ref{lemma1} we conclude that there is a closed circuit $\mathcal{C}(0)$ in $\mathcal{L}_h$ if $\lambda_f>3.65/r_r^2$. Meanwhile, the existence of $\mathcal{C}(0)$  assures $K_{\mathcal{L}_h}(0)<\infty$. Hence, $\lambda_f>3.65/r_r^2$ is the condition that guarantees that the ISG $\mathcal{I}(\Xi,\mathcal{E})$ does not percolate, which concludes the proof of Proposition \ref{Prop1}.
\end{IEEEproof}
 \end{prop}
 Before switching the discussion to the super-critical regime operation of the ISG, it is worth stating the following two important remarks.  
 \begin{remark}
 \label{remark:ind}
 It is important to note that the condition in \eqref{eq4} that enforces sub-critical regime operation of the ISG is independent of the IoT/CPS devices intensity $\lambda_r$. This is because the proof of Proposition~\ref{Prop1} is based on the collective ability of the secured zones to spatially quarantine malware infection within finite region. That is, the condition in \eqref{eq4} implies that the spatial firewalls are dense enough to construct continuous secured zones that surround any malware infection to safeguard the IoT/CPS from malware epidemic regardless of the IoT/CPS devices intensity.
 \end{remark}
 \begin{remark}
 	\label{remark:r}
 	The proof of Proposition \ref{Prop1} is based on the assumption that $r_r=r_f$. The case of $r_f > r_r$ also satisfies Definition~\ref{def_face} for the closed face. In fact, increasing $r_f$ expands the secured zone of $\mathcal{H}$. Hence, the expression for $\lambda_f$ that satisfies (\ref{eq6}) is also sufficient for no percolation when $ r_f\geq r_r$. Therefore, the proof of  Proposition \ref{Prop1} is also valid for the case of $r_f\geq r_r$.
 \end{remark}
 

 \subsection{Super-critical regime} \label{sec:super}
 
 This section shows that insufficient deployment of spatial firewalls leads to a super-critical regime operation for the ISG, which implies that the IoT/CPS network is at a risk of malware epidemics.  For a tractable analysis for the super-critical regime operation, we map the ISG to a square lattice as defined in the sequel.

 \textbf{Mapping to a Square Lattice:}
 Let $\mathcal{L}_s$ be a square lattice with side $s=\frac{r_r}{\sqrt{5}}$. The dual lattice $\mathcal{L}_s^d$ is a translated version of $\mathcal{L}_s$ with the translation magnitude ($\frac{s}{2}$,$\frac{s}{2}$). That is, $\mathcal{L}_s^d=\mathcal{L}_s+(\frac{s}{2},\frac{s}{2})$. Without loss of generality, it is assumed that one of the vertices of $\mathcal{L}_s^d$ is the origin. Let $\mathit{e}$ denote an edge common to two adjacent squares $S_1(\mathit{e})$ and $S_2(\mathit{e})$ in $\mathcal{L}_s
 $ and $\mathit{e}^d$ is the corresponding dual edge in $\mathcal{L}_s^d$. According to the spatial firewalls and IoT/CPS devices locations, the edge  $e$ can be either \textit{open} or \textit{closed} as defined below. 
 
 \begin{definition}[Open/closed edge]
 	\label{def_edge}
 	Let $\left\{\mathit{v_k} \right\}_{k=1}^4$ denote vertices of a rectangle formed by the union $S_1(\mathit{e}) \cup S_2(\mathit{e})$. Also, let $A(\mathit{e})$ be the smallest square containing circles $\left\{C(\mathit{v_k},r_f)\right\}_{k=1}^4$. where $C(a,r)$ denotes a circle of radius $r$ centered at $a$. Then an edge $e$ is defined to be open if i) each of $S_1(e)$ and $S_2(e)$ has at least one IoT/CPS device, and ii) there are no firewalls within $A(e)$. Otherwise, the edge is said to be closed. 
 \end{definition}

\begin{figure} [t!]
  	\centering
  	\includegraphics[scale=0.55]{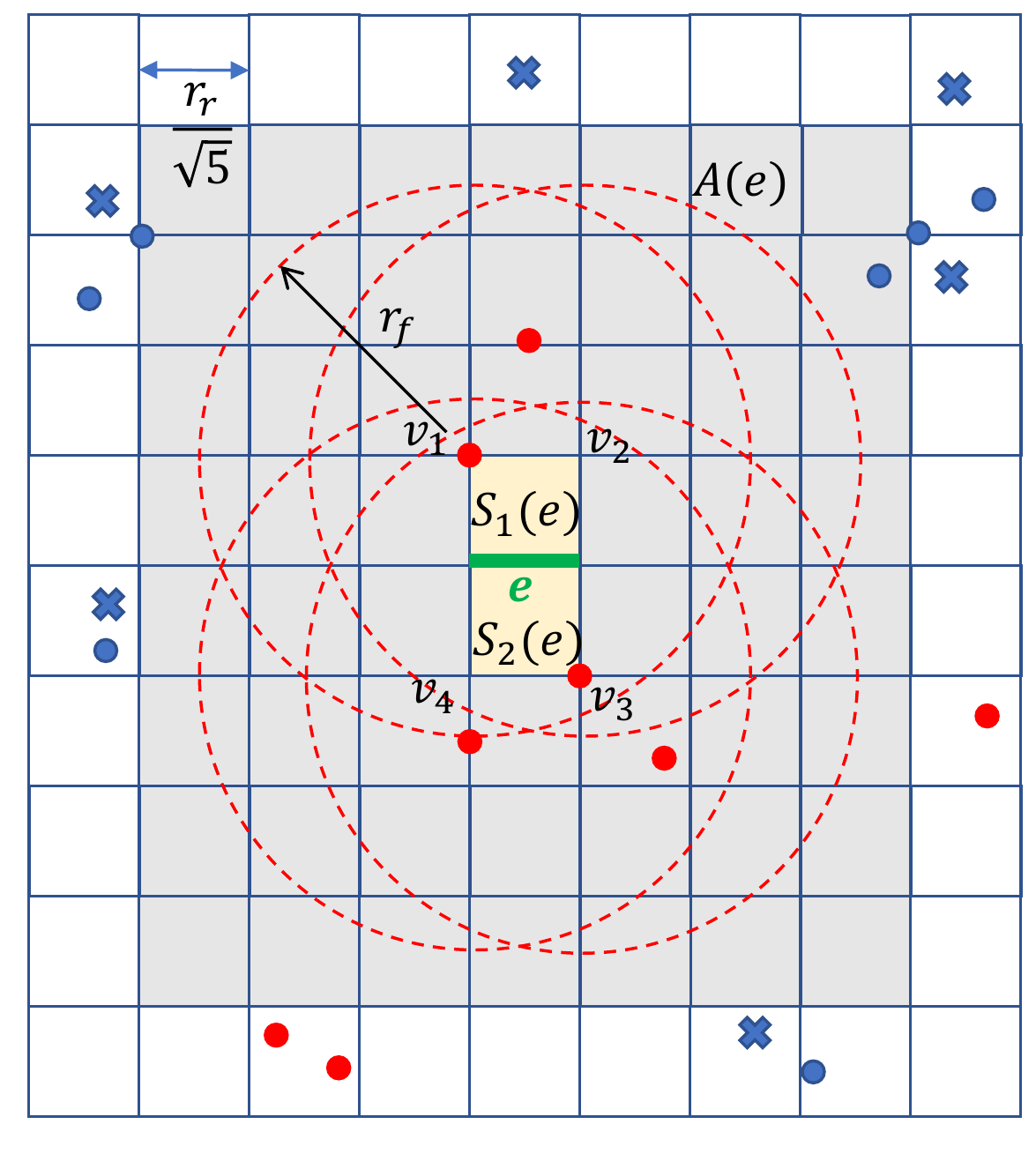}
  	\caption{An illustration of the open edge $e$ in the square lattice $\mathcal{L}_s.$, where red dots denote susceptible devices, blue dots denote protected devices, and crosses denote spatial firewalls.}
  	\label{Fig square}
  \end{figure}

  A pictorial illustration of the square lattice mapping with an open edge $e$ is shown in Fig~\ref{Fig square}. The rectangular lattice mapping is chosen to define a geographical region that contains a connected component of susceptible devices in the ISG $\mathcal{I}=\{\Xi,\mathcal{E}\}$. Since $A(e)$ is free from firewalls, then the region covered by $S_{1}(\mathit{e}) \cup S_{2}(\mathit{e})$ is located outside the secured zones of all firewalls. Furthermore, since each of  $S_{1}(\mathit{e})$ and $S_{2}(\mathit{e})$ has at least one IoT/CPS device, then the region $S_{1}(\mathit{e}) \cup S_{2}(\mathit{e})$ contains some vertices of $\Xi$. Lastly, since the largest distance (i.e., the diagonal) within the region $S_{1}(\mathit{e}) \cup S_{2}(\mathit{e})$ equals to $r_r$, all devices that are located within $S_{1}(\mathit{e}) \cup S_{2}(\mathit{e})$ are within the D2D range of each other. Hence, the region defined by $S_{1}(\mathit{e}) \cup S_{2}(\mathit{e})$ contains devices in $\Xi$ that are all connected to each other with edges in $\mathcal{E}$. {The connectivity within $S_{1}(\mathit{e}) \cup S_{2}(\mathit{e})$ (i.e., open edge in $\mathcal{L}_s$) is also represented via an open edge in the dual lattice $\mathcal{L}^d_s$. Hence, bond percolation on the square lattice $\mathcal{L}_s^d$ implies infinite connected component in $\mathcal{I}=\{\Xi,\mathcal{E}\}$.}  To study bond percolation of $\mathcal{L}_s^{d}$, we focus on the connected component that contains the origin. As mentioned before, there is no loss in generality to focus on the origin due to the stationarity of the PPP. The coupling between the square lattice $\mathcal{L}_s^d$ and the ISG $\mathcal{I}=\{\Xi,\mathcal{E}\}$ is formally stated in the following lemma.

  \begin{lemma}[Square lattice coupling]
  	\label{lemma:coupling}
  	Let $K_{\mathcal{L}_s^d}(0)$ denote a connected component in $\mathcal{L}_s^d$ containing the  origin. If $K_{\mathcal{L}_s^d}(0)$ is infinite, then $K_{\mathcal{I}}(0)$ is also infinite.
  	\begin{IEEEproof}
  		Let a path $\mathcal{P}_{\mathcal{L}_s^d}$ denote a  sequence  of connected  open edges in $\mathcal{L}_s^d$. Since there is one to one mapping between dual and prime edges, $\mathcal{P}_{\mathcal{L}_s^d}$ is uniquely associated with another path $\mathcal{P}_{\mathcal{L}_s}\in \mathcal{L}_s$, in which all edges are also open. Furthermore,  $\mathcal{P}_{\mathcal{L}_s}$ is associated with a unique sequence of $\left \{S_1(e_i),S_2(e_i)\right \}_{e_i\in \mathcal{P}_{\mathcal{L}_s}}$ pairs, where each pair is composed of single connected component within the ISG $\mathcal{I}$. Hence, an infinite-length path in $\mathcal{L}_s^d$ implies that there is an infinite sequence of connected susceptible devices that are members of the same connected component in $\mathcal{I}(\Xi,\mathcal{E})$. 
  	\end{IEEEproof}
  \end{lemma}
 
By virtue of Lemma \ref{lemma:coupling}, it is sufficient to characterize percolation in $\mathcal{L}_s^d$ to prove the super-critical regime of the ISG $\mathcal{I}(\Xi,\mathcal{E})$. However, before delving into the analysis, it is important to note the dependency between proximate edges, as stated in the following remark.
\begin{remark}[Edge dependencies]
	\label{correlation}
	Since an open edge $e$ ensures that firewalls are absent from the region $A(e)$, then the status of proximate edges are correlated. The status of two edges are independent if they do not share a common spatial region that requires the absence of spatial firewalls. {Hence, the smallest distance that ensures independent edges is $2s\lceil \frac{r_f}{s}\rceil$ horizontally and $2s\lceil \frac{r_f}{s}\rceil+2s$ vertically.}
\end{remark}

Now we are in position to study the super-critical regime of the ISG, which is characterized in the following proposition.

  \begin{prop}[Sufficient condition for non-zero percolation on ISG]
 	\label{prop2}
 	For given $r_r>0$ and $\lambda_r>0$, the ISG operates in the super-critical regime (i.e., $\theta_\mathcal{I}(\lambda_f,r_f,\lambda_r,r_r)>0$)  if 
 	\begin{equation}
 	\label{eqqq2}
 	\lambda_f< \frac{10}{N_A r_r^2}      \ln\left(\frac{1-\exp\left(-\frac{\lambda_r r_r^2}{5}\right)}{\sqrt{1-\beta}}\right), 
 	\end{equation}
 	where  $\beta=(\frac{11-2\sqrt{10}}{27})^N$,  $N=8ab-2a-6b+1$, $N_A=ab$, $a=2\lceil\frac{\sqrt{5}r_f}{r_r}\rceil+2$ and  $b=2\lceil\frac{\sqrt{5}r_f}{r_r}\rceil+1$.
 	\begin{IEEEproof}
	To prove the proposition, we characterize the conditions which ensure that the probability of no percolation is strictly less than one. Hence, the probability of the complement event (i.e., percolation) is strictly greater than zero. As mentioned in Lemma~\ref{lemma:coupling}, a finite $K_{\mathcal{L}_s^d}(0)$ implies no percolation on $\mathcal{L}_s^d$, which in turns implies sub-critical regime of the ISG. In the following, we find the conditions that ensures that $\mathbb{P}\{|K_{\mathcal{L}_s^d}(0)|<\infty\}<1$. Hence, such conditions also implies non-zero probability of the complement event  $\mathbb{P}\{|K_{\mathcal{L}_s^d}(0)|=\infty\}=(1-\mathbb{P}\{|K_{\mathcal{L}_s^d}(0)|<\infty\})>0$. Hence, the ISG has non-zero probability to operate in the super-critical regime.   
 	
 Let $\mathcal{P}_{\mathcal{L}_s}(n)$ denote a path of length $n$ edges $\left\{e_i\right\}_{i=1}^n \in \mathcal{L}_s$. From the coupling between the dual and primal square lattices, it can be inferred that $K_{\mathcal{L}_s^d}(0)$ is finite iff there is a closed circuit path in $\mathcal{L}_s$ around the origin. To account for the edge dependencies within the path $\mathcal{P}_{\mathcal{L}_s}(n)$, 
 {we recall from Definition~\ref{def_edge} that the edges $\mathit{e}_i$ and $\mathit{e}_j$ are independent if $(A(\mathit{e} _i) \cap A(\mathit{e}_j))=\emptyset$.  	{ Let $N$ denote the number of edges in $A_0(e)$ and let $S_I\subseteq \mathcal{P}_{\mathcal{L}_s}(n)$ denote the subset of all independent edges in $\mathcal{P}_{\mathcal{L}_s}(n)$.} Then, the set $S_I$ has a cardinality of at least $n/N$. The construction of $A_0(e)$ and the computation of $N$ are illustrated in Appendix \ref{Appendix}.}
 	
 	It is shown in \cite{Grimmett} that there are $4n3^{n-2}$ possible ways to construct a circuit of length $n$ around the origin. Therefore, the probability that a closed path exists around the origin  is expressed as  
 {\begin{align}
 		\label{eq:close}
 	 P_{c} &=  \sum_{n}4n3^{n-2} \mathbb{P}\{ \mathcal{P}_{\mathcal{L}_s}(n) \text{ is closed}\} \notag \\
 	 & \leq \sum_{n=1}^{\infty}4n3^{n-2}q^{\frac{n}{N}}   {=}\frac{4q^{\frac{1}{N}}}{3(1-3q^{\frac{1}{N}})^2},  
 		\end{align}}
 		\par
 		\noindent where $q\equiv \mathbb{P}\{e \text{ is closed}\}$ and the last equality in \eqref{eq:close} is obtained by treating the sum as a derivative of geometric series with respect to $q^{1/N}$. To ensure that $P_c$ is strictly less than 1, the following condition must be satisfied
 		\begin{equation}
 		\label{eq21}
 		q<(\frac{11-2\sqrt{10}}{27})^{N}.
 		\end{equation}
 		Based on Definition \ref{def_edge}, an explicit expression for $q$ can be found as follows  
 		\begin{align}
 	\label{eq 9}
 	q&=\! 1\!- \mathbb{P}\{ \Phi \cap S_1(e)\! \neq\! \emptyset \;\; \& \;\; \Phi \cap S_2(e)\! \neq\! \emptyset \;\;  \& \;\; \Psi \cap A(e)\! =\! \emptyset \} \notag \\
 	 &=1-(1-e^{-\lambda_rs^2})^2e^{-\lambda_f N_A s^2},
 	\end{align}
 	where $N_A$ is the number of squares covered by $A(e)$. From Fig. \ref{Fig square}, 
 		$N_A=(2\lceil\frac{\sqrt{5}r_f}{r_r}\rceil+2)\times(2\lceil\frac{\sqrt{5}r_f}{r_r}\rceil+1)$. The condition in \eqref{eq21} ensures that  $\mathbb{P}\{|K_{\mathcal{L}_s^d}(0)|<\infty\}<1$, which implies non-zero probability of percolation. Hence, substituting   (\ref{eq 9})
 		into (\ref{eq21}) and after some basic algebraic manipulations, we can finally get the result in (\ref{eqqq2}).
 	\end{IEEEproof}
 \end{prop}
 
\begin{remark}
	\label{remark:apparent}
	 Proposition~\ref{prop2} shows that the super-critical regime operation, which raises the risk  of malware epidemic requires both i) sufficiently dense IoT devices and ii) sufficiently sparse firewall deployment. More precisely, \eqref{eqqq2} shows an inverse relationship between $\lambda_r$ and $\lambda_f$ to allow long-range malware propagation. Hence, it may be apparent that higher intensity of IoT/CPS devices requires higher intensity of firewalls to spatially quarantine malware infections. However, this is only true up to the threshold shown in  Proposition~\ref{Prop1}. This is because the intensity shown in Proposition~\ref{Prop1} implies that the firewalls are dense enough such that the union of their secured zones form continuous circles in the spatial domain that surrounds and thwarts any emerging malware infection. Hence, the intensity of spatial firewalls shown in Proposition~\ref{Prop1} safeguards the IoT/CPS network from malware epidemics irrespective of the IoT/CPS devices intensity.
\end{remark}

 
 
  \subsection{Phase transition}\label{sec:mon}
 
This section shows that the percolation probability of the ISG exhibits a phase transition property in the intensity of firewalls $\lambda_f$. Hence, the percolation critical intensity of firewalls is the unique intensity that minimizes \eqref{object}. Such phase transition behavior is formally stated in the following theorem.

\begin{figure} [t!]
 
 \centering
\includegraphics[scale=0.6]{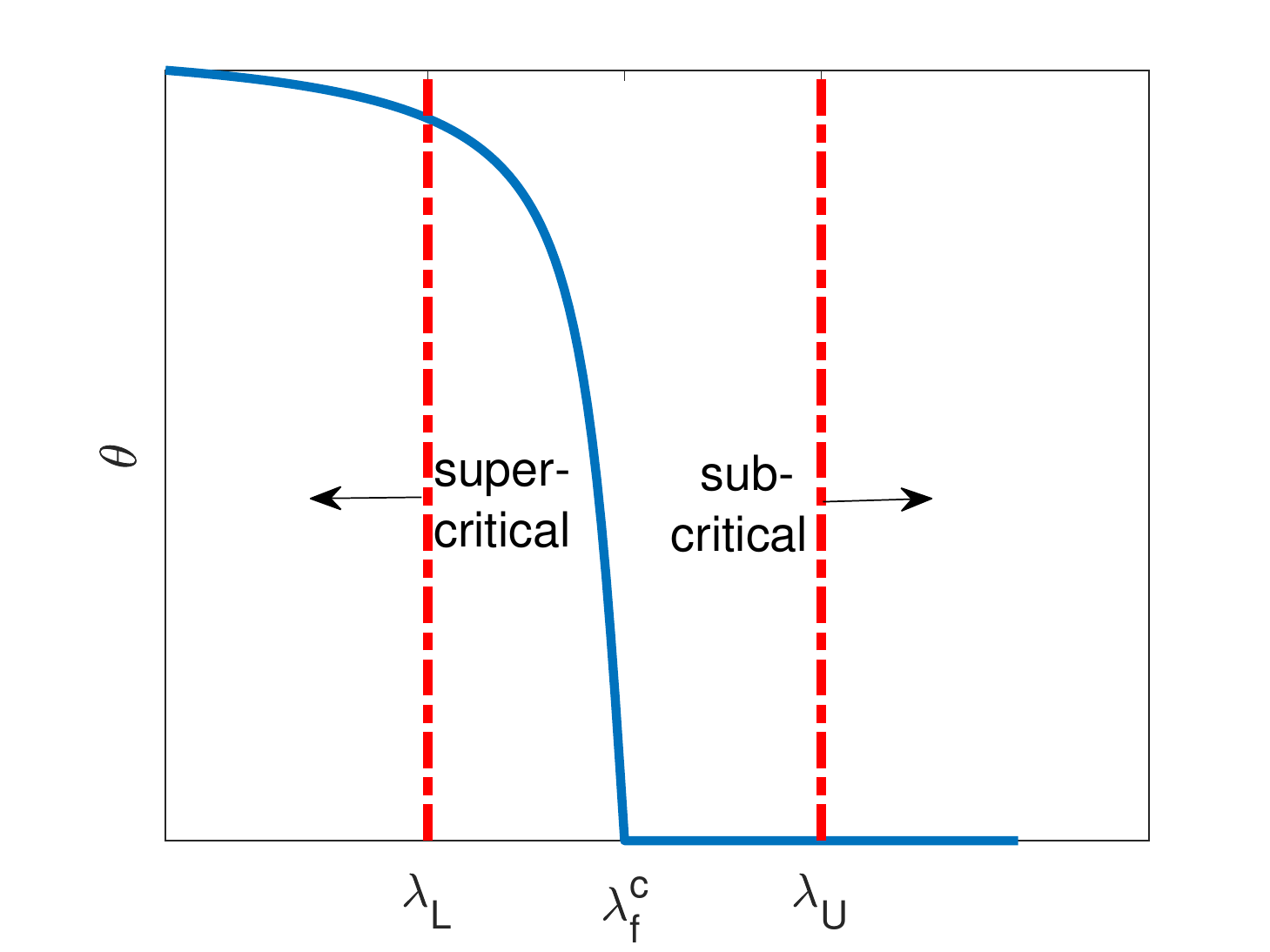}
\caption{Existence of critical density of firewalls.}
\label{continuity}
\end{figure}

\begin{theorem}[Phase transition]
	\label{Theorem: phase}
	Let $\theta_\mathcal{I}(\lambda_f,r_f,\lambda_r,r_r)$ denote the percolation probability of the ISG $\mathcal{I}(\Xi,\mathcal{E})$, then $\forall \lambda_r>0$ there exists a critical value $\lambda_f^c<\infty$ for the density of firewalls such that
	\label{th:2}
	\begin{equation}
	\label{eq2} 
	\begin{array}{ll}{\theta_\mathcal{I}(\lambda_f,r_f,\lambda_r,r_r)>0,} & {\text { for } \lambda_f <\lambda_f^c} \\ {\theta_\mathcal{I}(\lambda_f,r_f,\lambda_r,r_r)=0,} & {\text { for } \lambda_f>\lambda^c_f}\end{array}.
	\end{equation}
\end{theorem}
\begin{IEEEproof}
{	We start the proof by showing that the percolation probability  $\theta_\mathcal{I}(\lambda_f,r_f,\lambda_r,r_r)$ is non-increasing function in $\lambda_f$. Consider two sets of firewalls $\Psi_1$ and $\Psi_2$  with intensities $\lambda_{f_1}<\lambda_{f_2}$, respectively. Owing to the fact that both $\Psi_1$ and $\Psi_2$ are PPPs and that $\lambda_{f_1}<\lambda_{f_2}$, then $\Psi_1$ can be constructed by thinning $\Psi_2$ with probability $\frac{\lambda_{f_1}}{\lambda_{f_2}}$~\cite[Chapter~2]{martin_book}. As thinning implies random removal  of nodes, then $|\Psi_1\cap \mathcal{A}| \leq|\Psi_2\cap \mathcal{A}|$ for any $\mathcal{A} \in \mathbb{R}^2$. For the set of IoT devices $\Phi$, the set of protected devices is defined as $\Theta_i=\{x_i\in\Phi: \min\left\|x_i-\Psi_i\right\|<r_f\}$ for $i\in\{1,2\}$, and hence, $|\Theta_1\cap \mathcal{A}| \leq |\Theta_2\cap\mathcal{A}|$ for any $\mathcal{A} \in \mathbb{R}^2$. Now consider two ISGs $\mathcal{I}_1=(\Xi_1,\mathcal{E}_1)$ and $\mathcal{I}_2=(\Xi_2,\mathcal{E}_2)$ constructed with the same parameters $\lambda_r$, $r_r$ and $r_f$ but with the different sets of firewalls $\Psi_1$ and $\Psi_2$.	Since $\Xi_i=\Phi \setminus \Theta_i$ for $i\in\{1,2\}$, then $|\Xi_1\cap \mathcal{A}|\geq |\Xi_2\cap \mathcal{A}| $ for any $\mathcal{A} \in \mathbb{R}^2$. So, it is valid to state that  $K_{\mathcal{I}_1}(0)\supseteq K_{\mathcal{I}_2}(0)$. Therefore, the condition $0<\lambda_{f_1}<\lambda_{f_2}$ implies that $\theta_{\mathcal{I}}(\lambda_{f_1},r_f,\lambda_r,r_r) \geq \theta_{\mathcal{I}}(\lambda_{f_2},r_f,\lambda_r,r_r)$, and hence, $\theta_{\mathcal{I}}(\lambda_f)$ is a non-increasing function  of $\lambda_f$. }
	
	Recall that {$\theta_{\mathcal{I}}(\lambda_{f_1},r_f,\lambda_r,r_r)>0$} for $\lambda_f<\lambda_L$ as shown in Proposition~\ref{prop2}. Also, {$\theta_\mathcal{I}(\lambda_{f_1},r_f,\lambda_r,r_r)=0$} for $\lambda_f>\lambda_U$ as shown in Proposition~\ref{Prop1}. Since $\theta_\mathcal{I}(\lambda_{f_1},r_f,\lambda_r,r_r)$ is non-increasing in $\lambda_f$, there should be a critical value $\lambda_f^c$ that exhibit the phase transition indicated in \eqref{eq2} and depicted in Fig.\ref{continuity}. 		
\end{IEEEproof}

\section{Secured IoT/CPS Network Design}
\label{Sec:upper}

Section~\ref{Phase_transition} proves the concept of spatial firewalls. In particular, Theorem~\ref{Theorem: phase} shows that the phase transition critical intensity $\lambda_f^c$ is the minimum intensity of firewalls that safeguards the IoT/CPS from malware epidemics. Hence, $\lambda_f^c$ implies minimum deployment and licensing cost for spatial firewalls. However, as in the majority of continuum percolation models, there is no exact expression for $\lambda_f^c$. Hence, approximations and bounds are always sought. Proposition~\ref{Prop1} shows sufficient conditions for the firewalls intensity to safeguards IoT/CPS networks from malware epidemic. However, the sufficient condition of Proposition~\ref{Prop1} can be regarded as a loose upper-bound on $\lambda_f^c$ as it assumes a worst case scenario of $r_f=r_f$. Furthermore, Proposition~\ref{Prop1} restricts the vulnerable and secured regions to hexagonal shapes. Relaxing the assumptions of Proposition~\ref{Prop1}, the following theorem presents a tight upper-bound for $\lambda_f^c$, which provides an economical design of spatial firewalls. In addition, Theorem 2 gives extra parameter for manipulation of $\lambda_f^c$ upperbound. 

\begin{theorem}
\label{Theorem:upper}
Consider an IoT/CPS network with devices intensity $\lambda_r>0$ and D2D communications range $r_r>0$. To secure such IoT/CPS network, spatial firewalls with communication/detection range of $r_f\geq r_r$ are deployed. Then, the critical intensity of firewalls that safeguards such IoT/CPS network from malware epidemics is bounded by
{\begin{equation}
\label{eq:upperbound}
\lambda_{f}^{c} \leq \frac{\lambda_{c}(1)}{4  r_{f}^{2}-r_{r}^{2}},
\end{equation}}
where $\lambda_{c}(1)$ is given in Remark~\ref{remark22}.
\end{theorem}

\begin{figure} [t!]
	\centering
	\includegraphics[scale=0.8]{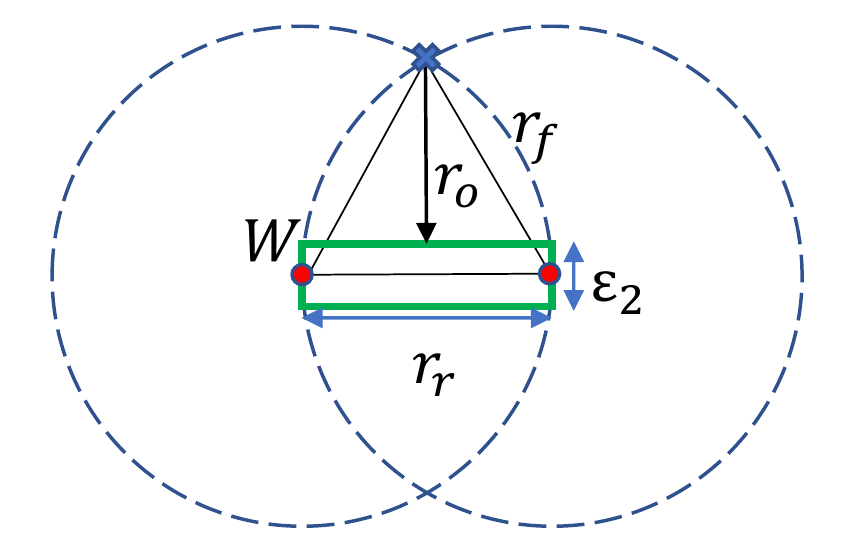}
	\caption{Worst case scenario for existence of infinite path. }
	\label{Fig 6}
\end{figure}

\begin{IEEEproof}
The construction of the ISG graph is based on the interaction between the IoT/CPS devices in $\Phi$ and the firewalls in $\Psi$. Particularly, the devices in the ISG $\Xi=\Phi\setminus\Theta$ are the IoT/CPS devices in $\Phi$ that exists outside the secured zones of the firewalls in $\Psi$. Hence, $\Xi$ and $\Psi$ can be treated as an overlay of two non-intersecting networks: a network of firewalls and a network of susceptible devices. Let us define the vacant space $\mathcal{V}=\{x \in\mathbb{R}^2: \min\left\|x-\Psi\right\|>r_f\}$ as all spatial regions in $\mathbb{R}^2$ that are not covered by the secured zones of the firewalls in $\Psi$. By virtue of the exclusive relation between $\mathcal{I}=\{\Xi,\mathcal{E}\}$ and $\Psi$, an infinite connected component in $\mathcal{I}$ necessitates an infinite vacant component within $\mathcal{V}$. Hence, to prove Theorem \ref{Theorem:upper}, we analyze the condition for
the existence of an infinite vacant component (i.e., infinite continuous space) in $\mathcal{V}$. Let us consider the worst-case arrangement shown in Fig. \ref{Fig 6}. The figure shows two IoT/CPS devices that are exactly $r_f$ away from the nearest firewall and $r_r$ away from each other. Such setup depicts the minimum vacant space $W$ that allows for malware propagation between two IoT/CPS devices in $\Phi$, where $r_o$ is the minimum distance from any firewall in $\Psi$ and $W$. Therefore, the minimum requirement for the existence of an infinite path in $\mathcal{I}$ corresponds to the case of having an infinite vacant component in the Poisson Boolean model \cite{Haengi_book} with the intensity $\lambda_f$ and radius $r_o\equiv \sqrt{r_f^2-\frac{r_r^2}{4}}-\frac{\epsilon_2}{2}$. {Following \cite{Swami} it can be shown that the critical intensity for coverage percolation of a Boolean model with secured zones of radius $r$ is }
\begin{equation}
\label{swam}
    \frac{\lambda_c(1)}{(2r)^2}.
\end{equation}
Substituting the value for $r_o$ in \eqref{swam}, we conclude that there is no infinite vacant component in $\mathcal{V}$ if 
\begin{equation}
\label{swam2}
   \lambda_f\geq\frac{\lambda_c(1)}{\left (2\sqrt{r_f^2-\frac{r_r^2}{4}}-\epsilon_2 \right )^2}.
\end{equation}
Hence, the firewall intensity in \eqref{swam2} prohibits percolation in  $\mathcal{I}$. Owing to the fact that the critical intensity is the minimum intensity of firewalls that prohibit percolation in $\mathcal{I}$ and taking the limit $\epsilon_2 \longrightarrow 0$, we finally get the upper-bound in (\ref{eq:upperbound}). 
\end{IEEEproof}

The results in  Theorem~\ref{Theorem:upper} shows that any firewall intensity equal to or above the threshold shown in \eqref{eq:upperbound} is ensured to safeguard the IoT/CPS networks from malware epidemics. Different from Theorem~\ref{Theorem: phase}, the threshold shown in \eqref{eq:upperbound} accounts for the larger communication/detection range of firewalls when compared to the IoT/CPS devices. Capitalizing on  Theorem~\ref{Theorem:upper}, it is possible to prohibit malware epidemics through the design of the communication range of the IoT/CPS devices as an alternative to deploying more spatial firewalls. However, it should be noted that the communication range of the IoT/CPS devices should ensure global network connectivity. The D2D communication range that ensures both global network connectivity and prohibits malware epidemics is stated in the following corollary.   

\begin{cor}\label{cor}
For a given $\lambda_f$, $r_f$, and $\lambda_r$, a safeguarded wide range connectivity for the IoT/CPS network can be ensured if the D2D range satisfies the following condition 
 \begin{equation}
 \label{d2drange}
     \sqrt{\frac{\lambda_c(1)}{\lambda_r}}\leq r_r \leq \sqrt{4r_f^2-\frac{\lambda_c(1)}{\lambda_f}}.
 \end{equation}
 \begin{IEEEproof}
  The corollary can be directly proved from Theorem \ref{Theorem:upper} and IoT/CPS connectivity condition in \eqref{eq:connectivity1}.
 \end{IEEEproof} 
\end{cor}

The conditions in \eqref{d2drange} ensures that the D2D communication range is sufficient for legitimate information dissemination within $G=\{\Phi,E\}$ but not for a malware epidemic outbreak on $\mathcal{I}(\Xi,\mathcal{E})$. Hence, Theorem~\ref{Theorem:upper} and Corollary~\ref{cor} allow alternative techniques for safeguarding IoT/CPS networks, namely, through  the firewalls intensity $\lambda_f$, the firewalls communication/detection range $r_f$, or the IoT/CPS D2D range $r_r$. Note that the ranges $r_f$ and $r_r$ can be controlled through the transmit powers and receivers sensitivities.       

 The results in Theorem \ref{Theorem:upper} and Corollary~\ref{cor} safeguard the IoT/CPS networks from malware epidemics. However,  Theorem \ref{Theorem:upper} and Corollary~\ref{cor} do not give insights about the percentages of susceptible and secured IoT devices. Hence, an alternative design objective is to ensure that a required percentage, denoted as $\delta_{\text{sec}}$, of the devices are secured as shown in the following corollary

 \begin{cor}\label{cor1}
The percentage of IoT/CPS devices that are protected from malware infiltration/infection is given by 
 \begin{equation}
 \label{eq:delta}
 \delta_{\text{sec}}= 1-\exp(-\pi \lambda_f  r_f^2).
 \end{equation}
 	\begin{IEEEproof}
 By definition, a device is protected from malware infection if it falls within the secured zones of firewalls. Hence, the corollary can be directly proved from void probability of the PPP.
 	\end{IEEEproof} 
 \end{cor}
{ Combining the results of Theorem~\ref{Theorem:upper} and Corollary~\ref{cor1}, it can be shown that using the critical intensity of firewalls in \eqref{eq:upperbound} corresponds to  securing the following critical percentage of IoT/CPS devices   
 \begin{equation}
\label{sec_dev1}
 \delta_{\text{sec}}^c = 1-\exp\left\{- \frac{\pi \lambda_{c}(1)}{4 -\left(\frac{r_{r}}{r_f}\right)^{2}}\right\}.
\end{equation}
Owing to the fact that  $r_f \geq r_r$ and using $\lambda_c(1)\approx1.44$, it can be shown that $\delta_{sec}^c$ is a decreasing function in $r_f$ which is bounded within the following range
 \begin{equation}
 \label{sec_dev2}
 0.67 \leq \delta_{\text{sec}}^c \leq 0.78,
 \end{equation}
 where the upper limit corresponds to $r_f=r_r$ and the lower limit corresponds to the $r_f \gg r_r$.
 Note that  $\delta_{\text{sec}}$ approaches the lower limit rapidly with increasing $r_f$, which appears within a quadratic term inside the exponential function. 
 \begin{remark}
 	The results in Theorem~\ref{Theorem:upper} and \eqref{sec_dev2} show that lower values of $r_f$ necessitate higher intensity of firewalls to safeguard the IoT/CPS network. That is, lower values of $r_f$ requires protecting at most $10\%$ more IoT/CPS devices to safeguard the IoT/CPS network from malware epidemics.   
 	\end{remark}
 The results in \eqref{sec_dev1} and \eqref{sec_dev2} also show the percentage of devices that are required to be secured by spatial firewalls to safeguard the  IoT/CPS networks from malware epidemics.}

\section{Numerical and Simulation Results}
\label{simulations}


The simulation of the IoT/CPS network is implemented in Matlab. In each simulation run, two independent PPPs with intensities $\lambda_f $ device/m$^{2}$ for firewalls and $\lambda_r$ device/m$^{2}$ for IoT/CPS devices are scattered over a square region of size $100\times 100$ m$^2$. The set of protected devices $\Theta$ and the set of susceptible devices $\Xi$ are first identified based on the distances between the IoT/CPS devices and firewalls. The ISG $\mathcal{I}(\Xi,\mathcal{E})$ is then constructed based on the D2D communication range among the susceptible devices in $\Xi$. Percolation is declared on $\mathcal{I}(\Xi,\mathcal{E})$ if it contains a connected component that spans the simulation region from left to right and from bottom to top. Percolation probability is then calculated by averaging over several realizations of the Monte-Carlo simulations. The unit of $\lambda_r$, $\lambda_f$, and $\lambda_f^c$ in all the figures is device per square meter.

\begin{figure}[t!]
	\centering
	\includegraphics[width=0.5\textwidth]{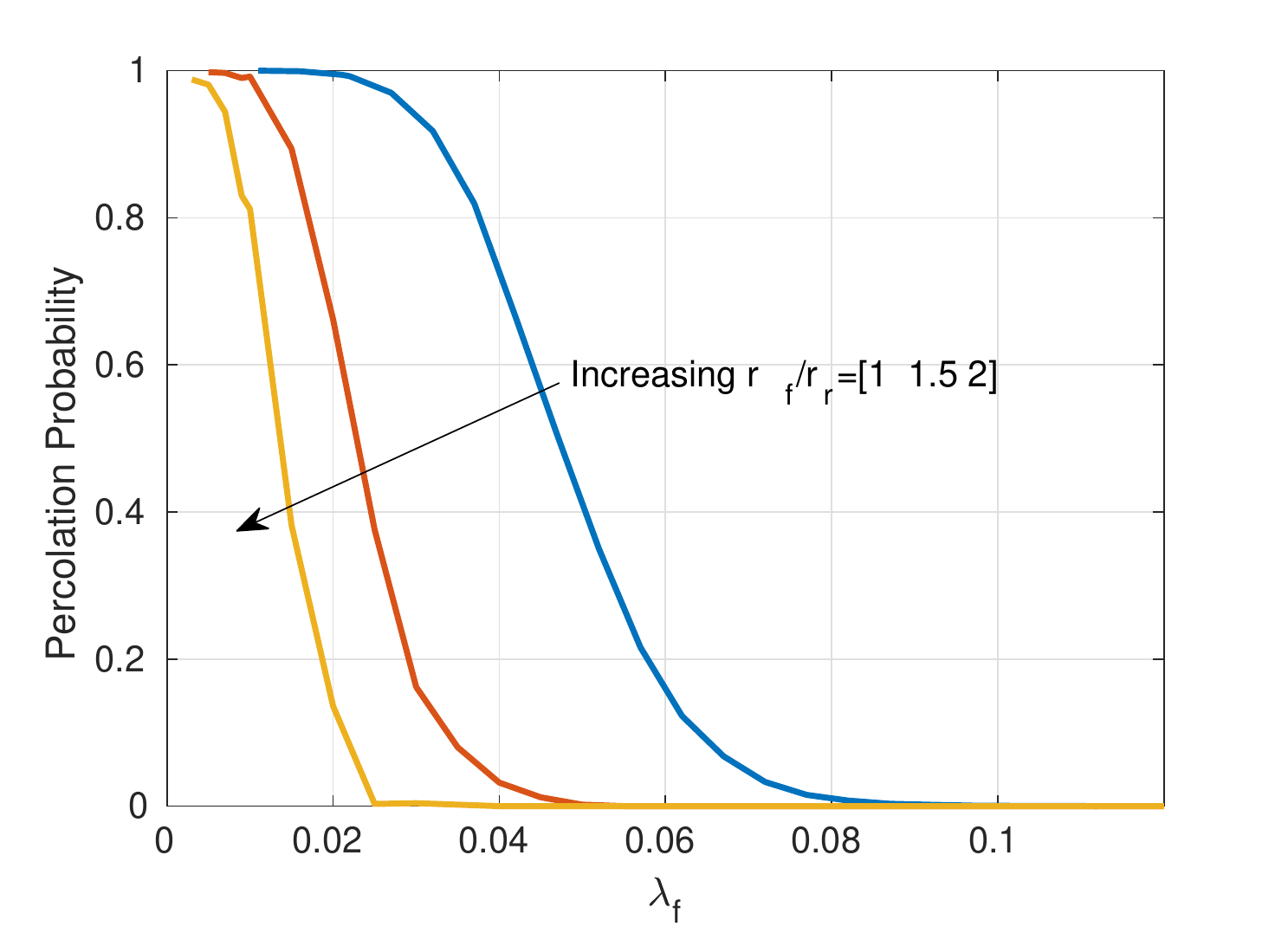}
	\caption{Phase transition.}
	\label{fig:2}
\end{figure}

Fig.\ref{fig:2} shows the percolation probability versus the intensity of firewalls for $r_r=2$ m and $\lambda_r=0.8$ device/m$^{2}$. Recall that the critical density, $\lambda_f^c$ appears at the point where percolation probability drops to 0 for the first time. The figure shows a phase transition in the percolation probability, which verifies Theorem \ref{Theorem: phase}. The figure also highlights the positive impact of firewall communication/detection range $r_f$. Increasing $r_f$ reduces the critical intensity that is required to safeguard the IoT/CPS network from malware epidemics.  

\begin{figure}
	\centering
\!\!\!\!\!\!	\begin{tikzpicture}[scale=0.5, transform shape,font=\large]
	\input{New_fig2}
	\end{tikzpicture}%
	\caption{The critical intensity of firewalls vs the intensity of IoT/CPS devices.}
	\label{fig:3}
\end{figure}
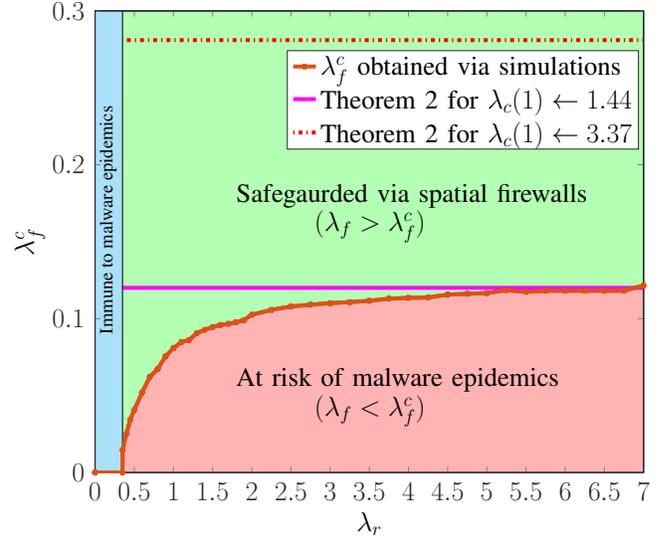

{ Fig. \ref{fig:3} plots the critical intensity of firewalls $\lambda_f^c$ (obtained via simulations) versus the intensity of IoT/CPS devices. The figure also depicts the upper-bounds in Theorem~\ref{Theorem:upper} for different values of $\lambda_c(1)$. The figure clearly shows the three regions of operation for the IoT/CPS network. The first region is characterized by small values of $\lambda_r$ such that the network lacks long range multi-hop wireless connectivity, and hence, is physically immune to malware epidemics. The second region is when the intensity of IoT/CPS is sufficiently high for global network connectivity while the intensity of firewalls is not sufficient (i.e., $\lambda_f<\lambda_f^c$), and hence, the network is at risk of malware epidemics. The last region is where sufficiently dense spatial firewalls are deployed (i.e., $\lambda_f\geq \lambda_f^c$), and hence, the network is safeguarded from malware epidemics. }

Fig. \ref{fig:3}  also shows that as the intensity of IoT/CPS devices increase, denser firewalls deployment is required to prohibit malware epidemics. However, the required intensity for firewalls saturates at the upper-bound indicated in Theorem~\ref{Theorem:upper}, which validates our analysis. That is, the figure confirms that there exists a finite intensity of firewalls $\lambda_f$ that safeguards the IoT/CPS network from malware epidemics regardless of the intensity of IoT/CPS devices. It is worth noting that the upper-bound in \eqref{eq:upperbound} depends on the approximate value of $\lambda_c(1)$. Hence, the figure also shows \eqref{eq:upperbound} when using the  upper-bound on $\lambda_c(1)$ provided in Remark~\ref{remark22}, which provides a safety margin against malware epidemics. { Note that operating with such safety margin increases the critical percentage of protected devices from the range shown in  \eqref{sec_dev2} to  $0.92 \leq \delta^c_{sec} \leq 0.97$ depending on the relative value of $r_f$ compared to $r_r$.}

\begin{figure}
    \centering
    \includegraphics[width=0.5
    \textwidth]{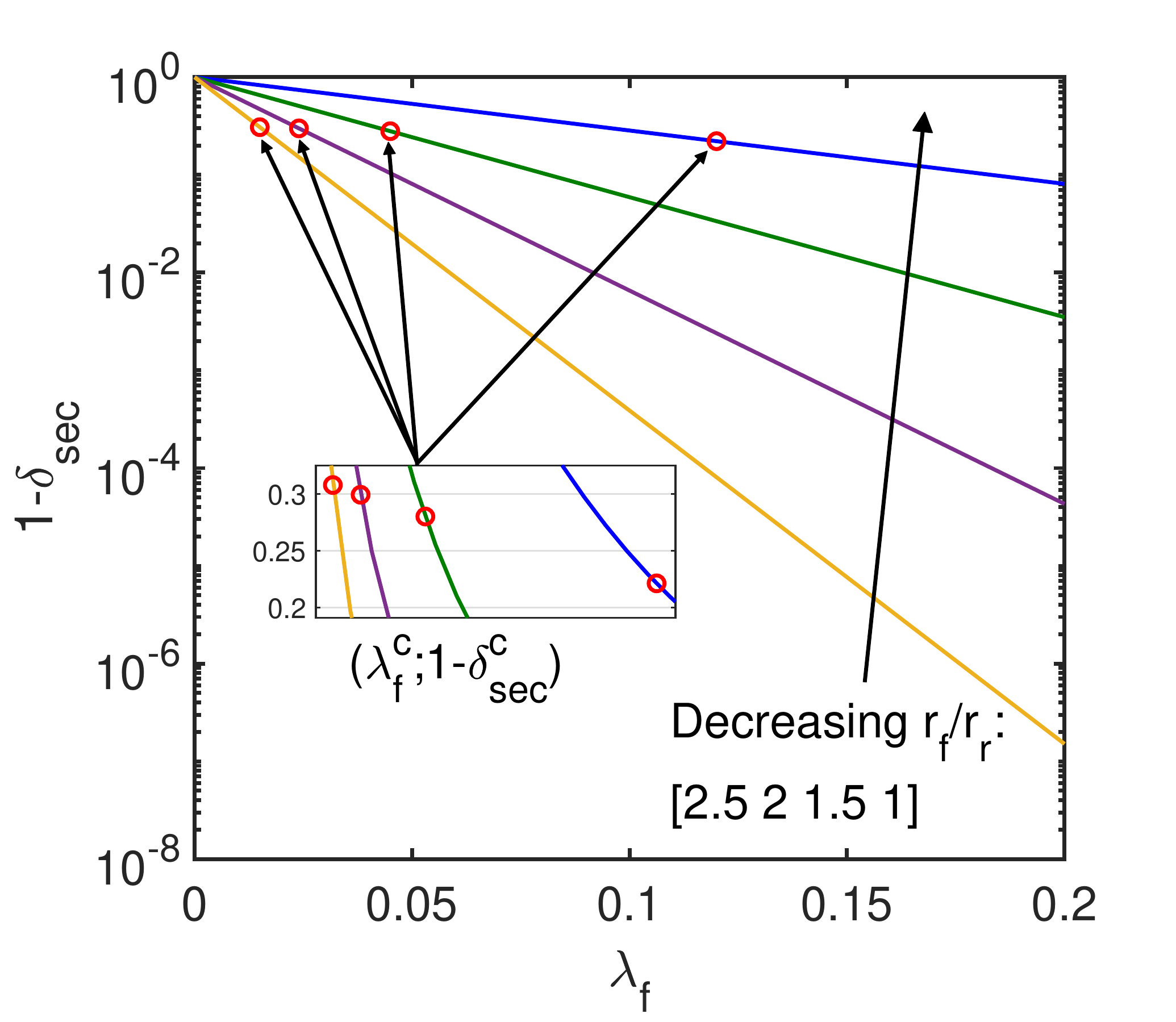}
    
    \caption{Percentage of susceptible devices vs different values of $\lambda_f$ (using $\lambda_c(1)\approx 1.44$).}
    \label{fig:delta}
\end{figure}

{Fig. \ref{fig:delta} shows the percentage of susceptible devices, i.e., ($1-\delta_{\rm sec}$), for different values of $\lambda_f$ and $r_f/r_r$. On the same figure, we also highlight the  critical percentage $1-\delta^c_{\rm sec}$, i.e., the complement of \eqref{sec_dev1}, which shows the percentage of susceptible devices when operating at the critical intensity $\lambda_f^c$ derived in Theorem~\ref{Theorem:upper}. First, the figure illustrates the range of $\delta_{sec}^c$ presented in \eqref{sec_dev2} and shows the fast convergence of $\delta_{sec}^c$ to the upper limit $0.78$. The figure also depicts the high impact of $r_f$ on the network design and performance. For instance, at $\lambda_f=0.1$ device/m$^2$ increasing $r_f$ to twice $r_r$ leads to more than 10-fold decrease in the percentage of susceptible devices. Furthermore, doubling $r_f$ also leads to around 4 times reduction in the critical intensity of firewalls required to safeguard the IoT/CPS devices. The figure also shows the costs, in terms of $\lambda_f$ and $r_f$, that are required to protect more devices beyond the required critical percentage  $\delta_{sec}^c$.

}

%
%

\section{Summary and Conclusions}
\label{conclusion}


{The Internet-of-things (IoT) is intrinsically vulnerable to large-scale malware attacks, where malware infiltrated to one device represents an infection threat to a large population of devices. To safeguard the IoT from malware epidemics, spatial firewalls are randomly deployed in the network to detect and thwart emerging malware infections. Each firewall imposes a secured zone, determined by its wireless connectivity, that protects its proximate devices form malware infection. Mapping the network to a random geometric graph and using tools from percolation theory, we develop a novel mathematical framework to assess and design spatial firewalls. In particular, we define the infection susceptible graph (ISG) to assess the risk of a malware outbreak (i.e., epidemic). We prove that the connectivity of the ISG exhibits a phase transition, where there exists a critical intensity of firewalls that prohibit the formation of a giant-connected component within the ISG. From the security perspective, the absence of a giant-connected component within the ISG eliminates the risk of long-range propagation of malware infection, and hence, safeguards the IoT from malware epidemics. }

To this end,  we present flexible design paradigm for the firewalls to safeguard the IoT. For instance, we find tight upper-bound for the critical intensity of spatial firewalls required to safeguard the IoT from malware epidemics. We also characterize the relative communications ranges of the firewalls and the IoT devices that allow secure global network connectivity. In addition, we specify the percentage of secured devices that corresponds to the critical intensity of spatial firewalls. The results show that the required density of spatial firewalls significantly decreases as we increase the relative communication range of the spatial firewalls when compared to the IoT devices. It is also shown that securing $67\%$ to $78\%$ of the IoT devices via firewalls is sufficient  to safeguard the IoT network from malware epidemics, where the required percentage of secured IoT devices decreases as the communication range of the spatial firewalls increases.

 \appendix
 \subsection{Maximum number of dependent edges}
 
  \label{Appendix}
  
\begin{figure} 
	
	\centering
	\includegraphics[scale=0.4]{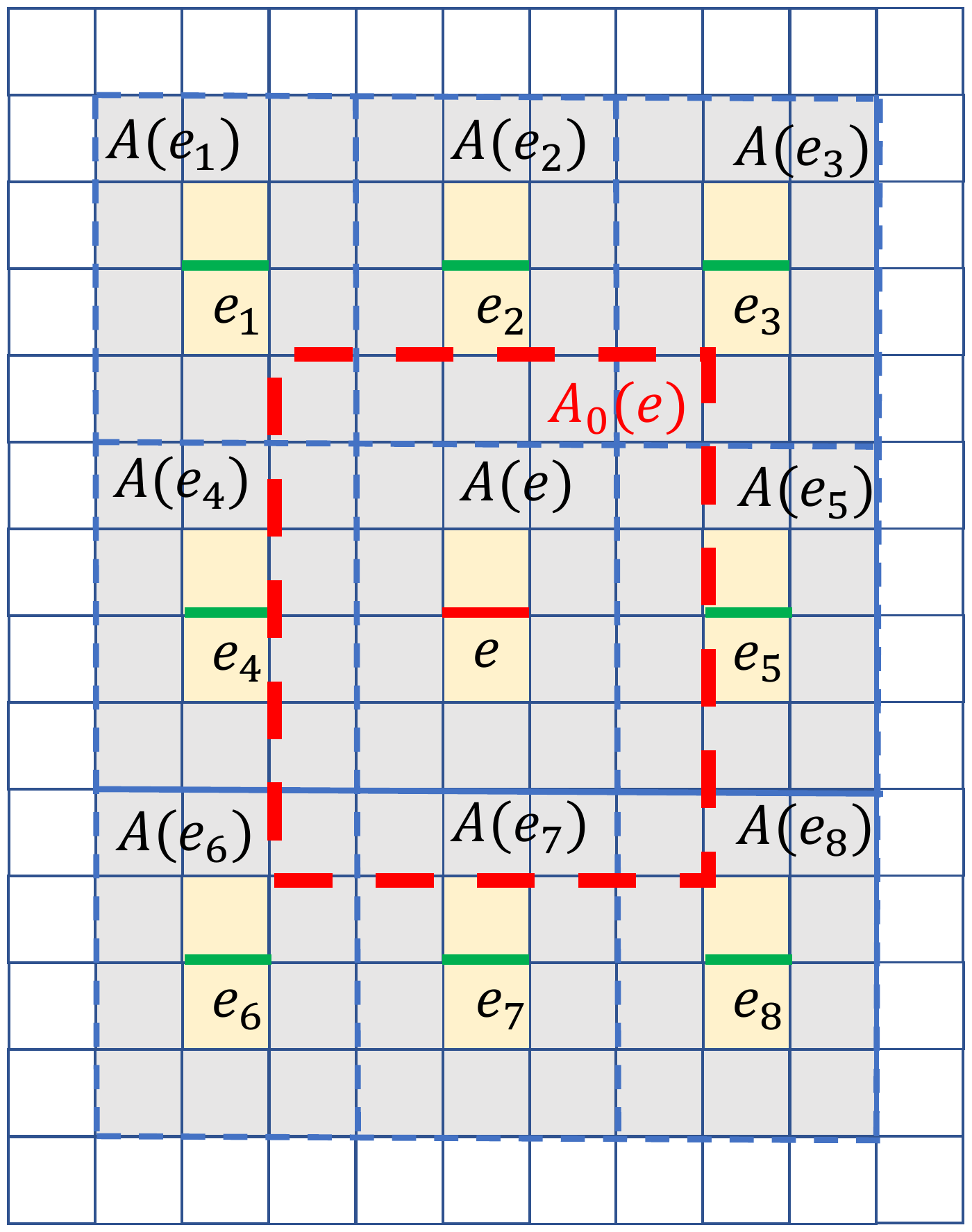}
	\caption{Example of edge dependency region $A(e)$: {$a=4$ and $b=3$}.}
	\label{fig:independence}
\end{figure}

 This appendix illustrates how to find general expression for the maximum number of edges  $N$ that are dependent of some arbitrary edge $e$. 
We assume that all edges in the lattice have dependency region $A(\cdot)$ of general size $a \times b$ squares. According to the definition of dependency region, any two edges $e_i, e_j$ are considered independent iff 
 $(A(\mathit{e} _i) \cap A(\mathit{e}_j))=\emptyset$. Therefore, let  $\left\{e_n\right\}^{8}_{n=1}$ be the 8 closest edges which dependency regions $\left\{A(e_i)\right\}^{8}_{i=1}$ do not overlap with $A(e)$. Next, we should construct maximal rectangle around $e$ such that created region $A_0(e)$ does not include $\left\{e_i\right\}^{8}_{i=1}$. Hence, we can be sure that $A_0(e)$ covers maximal number of edges that are dependent of $e$. For illustrative purposes, a toy example of edge dependency problem is given in Fig. \ref{fig:independence}. By 
 construction, the size of $A_0(e)$ is $(2a-2)\times (2b-1)$. Hence, the number of edges in $A_0(e)$ is
 \begin{equation}
 \label{eq:apendix}
 N=(2a-1)(2b-1)+(2a-2)2b=8ab-2a-6b+1.
 \end{equation}
{For the scenario in Proposition~\ref{prop2}, the dimensions of $A(e)$ are 
$$a=2\lceil\frac{\sqrt{5}r_f}{r_r}\rceil+2 \quad  \text{and} \quad b=2\lceil\frac{\sqrt{5}r_f}{r_r}\rceil+1.$$ 
Substituting these values in (\ref{eq:apendix}) leads to the final result of $N$. }

\bibliographystyle{IEEEtran}
\bibliography{Safegaurding_IoT_Final_Files.bbl}

\end{document}

%% file: notation.tex
\def\nb0{{\mathbf{0}}}
\def\nb1{{\mathbf{1}}}

\newtheorem{lemma}{Lemma}

\newtheorem{definition}{Definition}

\newtheorem{theorem}{Theorem}
\newtheorem{prop}{Proposition}
\newtheorem{cor}{Corollary}

\newtheorem{remark}{Remark}



%% file: defs_tikzpgf.tex
\usepackage{tikz}
\usepackage{pgfplotstable}
\usepackage{rotate}

\definecolor{rubblue}{cmyk}{1,0.5,0,0.6}
\definecolor{rubgreen}{cmyk}{0.5,0,1,0}
\definecolor{rubgray}{cmyk}{0.03,0.03,0.03,0.1}

\usepgfplotslibrary{units}

\usetikzlibrary{%
patterns,%
calc,%
fit,%
arrows,%
plotmarks,%
shadows,%
chains,%
shapes%
}

\tikzset{>=latex'} 
\tikzstyle{every picture}+=[remember picture] 
\pgfdeclarelayer{background}
\pgfdeclarelayer{foreground}
\pgfsetlayers{background,main,foreground}

\tikzstyle{blueblock}=[draw=rubblue, rectangle, thick, drop shadow, minimum width=20mm, minimum height=8mm,fill=rubblue!20, text width=20mm, text centered]
\tikzstyle{bluebox}=[draw=rubblue, rectangle, thick, drop shadow, minimum width=8mm, minimum height=8mm,fill=rubblue!20, text width=8mm, text centered]
\tikzstyle{greenblock}=[draw=rubgreen, rectangle, thick, drop shadow, minimum width=20mm, minimum height=8mm,fill=rubgreen!20, text width=20mm, text centered]
\tikzstyle{dot} = [draw, circle, minimum size=0.2pt,scale=0.3,fill=black,black]
\tikzstyle{smalldot} = [draw, circle, minimum size=0.1pt,scale=0.2,fill=black,black]
\tikzstyle{reddot}  =[draw,circle,minimum size=0.2pt,scale=0.8,fill=red,thin]
\tikzstyle{greendot}  =[draw,circle,minimum size=0.2pt,scale=0.8,fill=Green,thin]
\tikzstyle{bluedot}  =[draw,circle,minimum size=0.2pt,scale=0.8,fill=blue,thin]
\tikzstyle{whitedot}=[draw,circle,minimum size=0.2pt,scale=0.8,fill=white,thin]
\tikzstyle{blackdot} = [draw, circle, minimum size=0.2pt,scale=0.7,fill=black,black]
\tikzstyle{sum} = [drop shadow, draw=rubblue, thick, fill=rubblue!20, circle]
\tikzstyle{relay} = [blueblock, minimum width=5mm, minimum height=20mm, text width=5mm, rounded corners=2pt]
\tikzstyle{relay2} = [blueblock, minimum width=5mm, minimum height=15mm, text width=5mm, rounded corners=2pt]
\tikzstyle{relay3} = [blueblock, minimum width=5mm, minimum height=25mm, text width=5mm, rounded corners=2pt]
\tikzstyle{relay4} = [blueblock, minimum width=5mm, minimum height=10mm, text width=5mm, rounded corners=2pt]
\tikzstyle{relay5} = [blueblock, minimum width=5mm, minimum height=50mm, text width=5mm, rounded corners=2pt]
\tikzstyle{relay6} = [blueblock, minimum width=5mm, minimum height=5mm, text width=5mm, rounded corners=2pt]
\tikzstyle{circgreen} = [draw, circle, inner sep=2pt, fill=rubgreen, drop shadow, thick]
\tikzstyle{circwhite} = [draw, circle, inner sep=2pt, fill=white, drop shadow, thick]
\tikzstyle{circdashed} = [draw, dashed, circle, inner sep=2pt, fill=rubgray, drop shadow, thick]
\tikzstyle{vertbox} = [rectangle, draw=rubblue, thick, rotate=90, text centered, minimum width=16.5mm, minimum height=8mm, text width=16.5mm, inner sep=0pt, fill=rubblue!20, drop shadow]
\tikzstyle{vertboxb} = [rectangle, draw=rubblue, thick, rotate=90, text centered, minimum width=16.5mm, minimum height=8mm, text width=16.5mm, fill=rubblue!20, drop shadow]
\tikzstyle{vertboxshort} = [rectangle, draw=rubblue, thick, rotate=90, text centered, minimum width=10mm, minimum height=8mm, text width=10mm, inner sep=0pt, fill=rubblue!20, drop shadow]
\tikzstyle{smalldotgreen} = [draw=rubgreen, circle, minimum size=0.2pt,scale=0.8,fill=rubgreen!20]
\tikzstyle{antenna} = [regular polygon, regular polygon sides=3, draw, shape border rotate=180, minimum size=0.2pt, scale=0.3]

\tikzstyle{poly} = [regular polygon, regular polygon sides=6, shape aspect=0.5, minimum width=1.5cm, minimum height=0.35cm, draw, dashed]

\definecolor{cff9e00}{RGB}{255,158,0}
\definecolor{c4fff00}{RGB}{79,255,0}
\definecolor{cff0012}{RGB}{255,0,18}
\definecolor{c00c5ff}{RGB}{0,197,255}
\definecolor{c046f00}{RGB}{4,111,0}
\definecolor{c004b9d}{RGB}{0,75,157}

\newlength{\mylen}

%% file: New_fig2.tex
%
%
%
%
\definecolor{chromeyellow}{rgb}{1.0, 0.65, 0.0}
\definecolor{mycolor1}{rgb}{0.85,0.325,0.098}
\definecolor{mycolor2}{rgb}{1,0,1}

\definecolor{mygreentext}{RGB}{31,97,25} 
\definecolor{myredtext}{RGB}{207,21,24}
\definecolor{mybluetext}{RGB}{0,43,187} 
\definecolor{myblacktext}{RGB}{0,0,0} 
\definecolor{mygray}{RGB}{200,200,200} 
\definecolor{myorange}{RGB}{255, 178, 102}
\definecolor{mymagenta}{rgb}{0.78, 0.08, 0.52}
\definecolor{mymyblacktext}{rgb}{0, 0.74, 1}

\begin{axis}[%
view={0}{90},
width=5.74145833333333in,
height=4.83566666666667in,
scale only axis,
every outer x axis line/.append style={darkgray!60!black},
every x tick label/.append style={font=\color{darkgray!60!black}},
xmin=0, xmax=7,
every outer y axis line/.append style={darkgray!60!black},
every y tick label/.append style={font=\color{darkgray!60!black}},
ymin=0, ymax=0.3,
ytick={0,{0.05},0.1,0.15,0.2,0.25,0.3},
xtick=\empty, ytick=\empty,
legend style={at={(1,1)}, anchor=south west, legend columns=1, legend cell align=left, align=left, draw=white!15!black}]


\addplot [
fill=green!30!white,
]
coordinates{
	(0,0)(0,0.3)(7,0.3)(7,0)
};

\addplot [
fill=red!30!white,
forget plot
]
coordinates{
	(0.3499,0)(0.35,0.0145)(0.4,0.02518235294)(0.45,0.03441764706)(0.5,0.0406047619)(0.6,0.05198571429)(0.7,0.06225384615)(0.8,0.06732307692)(0.9,0.0755)(1,0.08077931034)(1.1,0.08468214286)(1.2,0.0861137931)(1.3,0.090625)(1.4,0.0927137931)(1.5,0.09449655172)(1.6,0.09588)(1.7,0.096612)(1.8,0.097676)(1.9,0.099068)(2,0.1024916667)(2.25,0.1056625)(2.5,0.1078833333)(2.75,0.1090285714)(3,0.1100631579)(3.25,0.11072)(3.5,0.11155)(3.75,0.11297)(4,0.1134615385)(4.25,0.1137416667)(4.5,0.1156)(4.75,0.1161090909)(5,0.1165833333)(5.25,0.1185222222)(5.5,0.1174)(5.75,0.1181)(6,0.1181)(6.25,0.1181)(6.5,0.1181)(6.75,0.1181)(7,0.1215)(7,0)
};

\end{axis}

\begin{axis}[%
view={0}{90},
width=5.74145833333333in,
height=4.83566666666667in,
scale only axis,
every outer x axis line/.append style={darkgray!60!black},
every x tick label/.append style={font=\color{darkgray!60!black}},
xmin=0, xmax=7,
every outer y axis line/.append style={darkgray!60!black},
every y tick label/.append style={font=\color{darkgray!60!black}},
ymin=0, ymax=0.3,
ytick={0,0.1,0.2,0.3},
xtick=\empty, ytick=\empty,
legend style={at={(0.35,0.70)}, font=\huge,style={row sep=0.15cm}, anchor=south west, legend columns=1, legend cell align=left, align=left, draw=white!15!black}]

\addplot [
color=mycolor1,
solid,
line width=3.0pt,
mark=asterisk,
mark options={solid}
]
coordinates{
 (0,0)(0.3499,0)(0.35,0.0145)(0.4,0.02518235294)(0.45,0.03441764706)(0.5,0.0406047619)(0.6,0.05198571429)(0.7,0.06225384615)(0.8,0.06732307692)(0.9,0.0755)(1,0.08077931034)(1.1,0.08468214286)(1.2,0.0861137931)(1.3,0.090625)(1.4,0.0927137931)(1.5,0.09449655172)(1.6,0.09588)(1.7,0.096612)(1.8,0.097676)(1.9,0.099068)(2,0.1024916667)(2.25,0.1056625)(2.5,0.1078833333)(2.75,0.1090285714)(3,0.1100631579)(3.25,0.11072)(3.5,0.11155)(3.75,0.11297)(4,0.1134615385)(4.25,0.1137416667)(4.5,0.1156)(4.75,0.1161090909)(5,0.1165833333)(5.25,0.1185222222)(5.5,0.1174)(5.75,0.1181)(6,0.1181)(6.25,0.1181)(6.5,0.1181)(6.75,0.1181)(7,0.1215) 
};
\addlegendentry{{$\lambda_f^c$ obtained via simulations}}

\addplot [
color=mycolor2,
fill=cyan,
fill opacity=0.5,
solid,
line width=3.0pt,
]
coordinates{
 (0,0.12)(0.3499,0.12)(0.35,0.12)(0.4,0.12)(0.45,0.12)(0.5,0.12)(0.6,0.12)(0.7,0.12)(0.8,0.12)(0.9,0.12)(1,0.12)(1.1,0.12)(1.2,0.12)(1.3,0.12)(1.4,0.12)(1.5,0.12)(1.6,0.12)(1.7,0.12)(1.8,0.12)(1.9,0.12)(2,0.12)(2.25,0.12)(2.5,0.12)(2.75,0.12)(3,0.12)(3.25,0.12)(3.5,0.12)(3.75,0.12)(4,0.12)(4.25,0.12)(4.5,0.12)(4.75,0.12)(5,0.12)(5.25,0.12)(5.5,0.12)(5.75,0.12)(6,0.12)(6.25,0.12)(6.5,0.12)(6.75,0.12)(7,0.12) 
};
\addlegendentry{{Theorem 2 for $\lambda_c(1) \leftarrow 1.44$}}

\addplot [
color=red,
dash pattern=on 1pt off 3pt on 3pt off 3pt,
line width=3.0pt
]
coordinates{
 (0,0.281)(0.3499,0.281)(0.35,0.281)(0.4,0.281)(0.45,0.281)(0.5,0.281)(0.6,0.281)(0.7,0.281)(0.8,0.281)(0.9,0.281)(1,0.281)(1.1,0.281)(1.2,0.281)(1.3,0.281)(1.4,0.281)(1.5,0.281)(1.6,0.281)(1.7,0.281)(1.8,0.281)(1.9,0.281)(2,0.281)(2.25,0.281)(2.5,0.281)(2.75,0.281)(3,0.281)(3.25,0.281)(3.5,0.281)(3.75,0.281)(4,0.281)(4.25,0.281)(4.5,0.281)(4.75,0.281)(5,0.281)(5.25,0.281)(5.5,0.281)(5.75,0.281)(6,0.281)(6.25,0.281)(6.5,0.281)(6.75,0.281)(7,0.281) 
};
\addlegendentry{{Theorem 2 for $\lambda_c(1) \leftarrow 3.37$}}

\addplot [
fill=cyan!30!white,
forget plot
]
coordinates{
	(0,0)(0,0.3)(0.3499,0.3)(0.3499,0)
};

\node[right, inner sep=0mm, rotate=90, text=black]
at (axis cs:0.167775474114275, 0.09) {\Large{Immune to malware epidemics}};
\node[right, inner sep=0mm, rotate=0, text=black]
at (axis cs:1.8,0.18) {\huge{Safegaurded via spatial firewalls}};
\node[right, inner sep=0mm, rotate=0, text=black]
at (axis cs:2.8, 0.16) {\huge{$(\lambda_f > \lambda_f^c)$}};
\node[right, inner sep=0mm, rotate=0, text=black]
at (axis cs:1.8, 0.06) {\huge{At risk of malware epidemics}};
\node[right, inner sep=0mm, rotate=0, text=black]
at (axis cs:2.8, 0.04) {\huge{$(\lambda_f < \lambda_f^c)$}};
\end{axis}

\begin{axis}[%
view={0}{90},
width=5.74145833333333in,
height=4.83566666666667in,
scale only axis,
every outer x axis line/.append style={font=\huge \color{darkgray!60!black}},
every x tick label/.append style={ font=\huge \color{darkgray!60!black}, yshift=-0.15cm},
xmin=0, xmax=7,
xlabel={\huge{$\lambda_r$}},
xlabel style={yshift=-0.4cm},
xtick={0,0.5,1,1.5,2,2.5,3,3.5,4,4.5,5,5.5,6,6.5,7},
every outer y axis line/.append style={darkgray!60!black},
every y tick label/.append style={font=\huge \color{darkgray!60!black}, xshift=-0.15cm},
ymin=0, ymax=0.3,
ytick={0,0.1,0.2,0.3},
ylabel={\huge{$\lambda^c_f$}},
ylabel style={yshift=0.5cm},
legend style={at={(0.58,0.75)}, font=\huge, style={row sep=0.5cm}, anchor=south west, legend columns=1, legend cell align=left, align=left, draw=white!15!black}]

\addplot [
color=mycolor1,
solid,
line width=3.0pt,
mark=asterisk,
mark options={solid}
]
coordinates{
	(0,0)(0.3499,0)(0.35,0.0145)(0.4,0.02518235294)(0.45,0.03441764706)(0.5,0.0406047619)(0.6,0.05198571429)(0.7,0.06225384615)(0.8,0.06732307692)(0.9,0.0755)(1,0.08077931034)(1.1,0.08468214286)(1.2,0.0861137931)(1.3,0.090625)(1.4,0.0927137931)(1.5,0.09449655172)(1.6,0.09588)(1.7,0.096612)(1.8,0.097676)(1.9,0.099068)(2,0.1024916667)(2.25,0.1056625)(2.5,0.1078833333)(2.75,0.1090285714)(3,0.1100631579)(3.25,0.11072)(3.5,0.11155)(3.75,0.11297)(4,0.1134615385)(4.25,0.1137416667)(4.5,0.1156)(4.75,0.1161090909)(5,0.1165833333)(5.25,0.1185222222)(5.5,0.1174)(5.75,0.1181)(6,0.1181)(6.25,0.1181)(6.5,0.1181)(6.75,0.1181)(7,0.1215) 
};
\end{axis}